\documentclass{JHEP3}

\usepackage{amsfonts,amsmath,amssymb,amsthm,axodraw}
\usepackage{latexsym}
\usepackage{graphicx}
\usepackage{cite}

\numberwithin{equation}{section}

\def\stamp{--- {\bf \today} --- {\bf \jobname.tex}}

\def\A#1#2{\la#1#2\ra}

\def\AB#1#2#3{\la#1|#2|#3]}

\def\nn{\nonumber}
\def\sg(#1){\textrm{sign}(#1)}
\def\cI{\mathcal{I}}
\def\cP{\mathcal{P}}
\def\I{\textrm{I}}
\def\J{\textrm{J}}
\def\m{\textrm{mass}}
\def\ol{\textrm{one-loop}}
\def\tree{\textrm{tree}}
\def\cN{\mathcal{N}}
\def\cA{\mathcal{A}}
\def\cM{\mathcal{M}}
\def\kf#1{K^\flat_#1}
\def\skf#1{\not\!K^\flat_#1}
\def\ra{\rangle}
\def\la{\langle}

\def\XX{\chi}
\def\AB#1#2#3{\langle #1 | #2 | #3]}
\def\e{\epsilon}
\def\bl{\bar \ell}
\def\A#1#2{\la#1#2\ra}
\def\wh#1{\widehat{#1}}

\def\sla#1{\not\!{#1}}

\def\an[#1,#2]{\left\langle#1\,#2\right\rangle}
\def\aq[#1,#2,#3]{\left\langle#1|#2|#3\right]}
\def\qa[#1,#2,#3]{\left[#1|#2|#3\right\rangle}
\def\sq[#1,#2]{\left[#1\,#2\right]}
\def\spa#1.#2{\left\langle#1\,#2\right\rangle}
\def\spab#1.#2.#3{\left\langle#1|#2|#3\right]}
\def\spb#1.#2{\left[#1\,#2\right]}
\def\lor#1.#2{\left(#1\,#2\right)}

\preprint{DESY-08-168; IPhT-T-08-156; IHES-P-08-54}

\title{Simplicity in the Structure of QED and Gravity Amplitudes}

\author{Simon Badger\\
Deutches Elektronen-Synchrotron DESY,\\
Platanenallee, 6, D-15738 Zeuthen, Germany\\
\email{simon.badger@desy.de}}
\author{N.~E.~J.~Bjerrum-Bohr\\
School of Natural Sciences,\\
Institute for Advanced Study,\\
Einstein Drive, Princeton,\\
NJ 08540, USA\\
\email{bjbohr@ias.edu}}
\author{Pierre Vanhove\\
Institut des Hautes Etudes Scientifiques\\
Le-Bois-Marie\\
F-91440 Bures-sur-Yvette, France\\
\textrm{and}\\
Institut de Physique Th{\'e}orique,\\
CEA, IPhT, F-91191 Gif-sur-Yvette, France\\
CNRS, URA 2306, F-91191 Gif-sur-Yvette, France\\
\email{pierre.vanhove@cea.fr}}

\abstract{We investigate generic properties of one-loop amplitudes
in unordered gauge theories in four dimensions. For such theories
the organisation of amplitudes in manifestly crossing symmetric
expressions poses restrictions on their structure and results in
remarkable cancellations. We show that one-loop multi-photon
amplitudes in QED with at least  eight external  photons  are  given
only by scalar box integral functions. This QED `no-triangle'
property is true for all helicity configurations and has
similarities to the `no-triangle' property found in the case of
maximal ${\cal N} = 8$ supergravity. Results are derived both via a
world-line formalism as well as using on-shell unitarity methods. We
show that the simple structure of the loop amplitude originates from
the extremely good BCFW scaling behaviour of the QED
tree-amplitude.}


\newpage

\begin{document}

\section{Introduction}
Powerful methods based on on-shell unitarity have led to the discovery
of remarkable simplifications  in analytic expressions of perturbative
amplitudes      in       gauge      theory~\cite{Bern:2007dw}      and
gravity~\cite{Bern:1998sv,Bern:2005bb,BjerrumBohr:2005xx,
  BjerrumBohr:2005jr,BjerrumBohr:2006yw,Bern:2007xj}.    At   one-loop
order in four dimensions all $n$-point amplitudes can be expanded in a
set of basis functions consisting  of scalar box, triangle, and bubble
integrals        and        possibly        rational        polynomial
functions~\cite{Bern:1993kr,Bern:1994zx}. This is a consequence of the
kinematic  restrictions  induced   by  working  with  four-dimensional
momenta  and  the observation  that  amplitude expressions  containing
tensor integrals can be  reduced to scalar integrals, although through
extensive             and             cumbersome             algebraic
manipulations~\cite{intred,Campbell:1996zw,Bern:1993kr,Denner:2005nn}.
Generalisations to higher loop amplitudes are more complicated since a
generic  basis of  integral functions  is not  known for  an arbitrary
number of legs. This persists even in relatively simple examples, such
as   in the planar  limit of $\cN=4$ super  Yang-Mills where
dual conformal symmetry restricts the form of amplitudes.  Crossing
symmetry  in   colourless  theories  require  that   both  planar  and
non-planar  integrals are  present in  the amplitude.  This  makes the
construction  of  a basis  of  integral  functions  that captures  the
ultra-violet and  infra-red behaviour  of higher loop  amplitudes more
elusive~\cite{Bern:2008ap,Cachazo:2008vp,
  Cachazo:2008hp,Spradlin:2008uu}.    Nevertheless,  constraints  from
supersymmetry~\cite{Berkovits:2006vc},          string  theory       duality
arguments~\cite{Green:2006gt,Green:2006yu}   and   the  field   theory
computations   of~\cite{Bern:2006kd,Bern:2007hh,Bern:2008pv}    have
explicitly  shown that multi-loop  amplitudes in  $\cN=8$ supergravity
have  much simpler  forms  than  one would  expect  from a  Lagrangian
perspective.

In the case of colourless gauge theories the summation over all
orderings of external legs (this includes both planar and
non-planar contributions) leads to extra important cancellations in the
amplitude. At one-loop order such cancellations lead to the `no-triangle'
property~\cite{BjerrumBohr:2005xx,BjerrumBohr:2006yw,BjerrumBohr:2008ji,BjerrumBohr:2008vc,ArkaniHamed:2008gz}
of $\cN=8$ supergravity. We will show in this paper that for one-loop
multi-photon amplitudes with $n\geq 8$ legs we have a similar
`no-triangle' property.

Recently, it has been shown in~\cite{BjerrumBohr:2008ji} that the
world-line formalism (a.k.a. the `string based method' for field
theory amplitude computations~\cite{Bern:1991aq,Bern:1991an,Strassler:1992zr,Bern:1992ad,Schubert:2001he})
is particularly well suited for exhibiting the cancellations
coming from the summation over the various orderings of colourless
gauge theories at one-loop order. The higher-loop extension of this
formalism~\cite{Schmidt:1994zj,Roland:1996im,Schubert:2001he,Dai:2006vj}
presents a possible dimension independent framework for investigations
of the improved ultra-violet behaviour of maximal ${\cal N} = 8$
supergravity~\cite{BjerrumBohr:2006yw,Green:2006gt,Bern:2006kd,Green:2006yu,Berkovits:2006vc,Bern:2007hh,Bern:2007xj,Green:2008bf}.
The power of the world-line formalism has been demonstrated in the recent
study~\cite{Green:2008bf}
of the two-loop supergravity four-graviton $\cN=8$ amplitude
given in ref.~\cite{Bern:1998ug}.

The unordered cancellations featured in gravity theories are also
present in one-loop multi-photon amplitudes in QED and Super-QED but
in a simpler framework. One-loop multi-photon amplitudes with $n$
external legs have {\it na\" \i vely} $n$ powers of loop momenta.
This indicates that under $n$ steps of Passarino-Veltman
reductions~\cite{intred,Bern:1993kr,Campbell:1996zw,Denner:2005nn}
the amplitude would, {\it a priori}, contain scalar box, triangle
and bubble integrals and may also contain rational polynomial
(non-logarithmic) contributions. However, explicit computations
show that the true structure is somewhat simpler.

For the case of multi-photon massless QED amplitudes we can explain
the discrepancy between the {\it na\" \i ve} power counting and
explicit results using reduction formul\ae\, for unordered
amplitudes~\cite{BjerrumBohr:2008ji}, that were derived using the
world-line formalism. We will directly show that these reduction
formul\ae\ imply that the four-photon one-loop amplitude reduces down
to scalar box triangle and bubble integral functions together with
rational terms. For the six-photon amplitude that it reduces to
box and three-mass scalar triangle integral functions.
For multi-photon amplitudes beyond six-point we have a reduction
down to scalar box integral functions only. This box structure only
of the amplitude for $n\geq8$ external photons is true for all
helicity configurations but the precise  expansion  of  the
amplitude in terms of scalar  box  integral functions depends on the
choice of helicity for the external states.
Our observations are in complete correspondence with the recent
direct evaluation of the one-loop six-photon amplitude
in~\cite{Binoth:2007ca,Bernicot:2007hs}. An earlier evaluation of
the multi-photon MHV amplitude by Mahlon~\cite{Mahlon:1993fe} showed
that in this case the amplitude contains only massive box integral
contributions. This sheds further light on how considering
expressions for unordered amplitudes with full crossing symmetry
leads to a surprising simplicity for amplitudes.

The structure of the paper is as follows. In section~\ref{sec:stringbased}
we review  in details the recent results for the reductions of unordered
integral functions at one-loop. On general grounds we will then investigate
via a string-based formalism how reduction  formul\ae\ can be  induced
by invariance of amplitudes under gauge
transformations~\cite{BjerrumBohr:2008vc,BjerrumBohr:2008ji,BjerrumBohr:2008dp}.
In section~\ref{sec:RedFor} we will re-derive the reduction formul\ae\ in
ref.~\cite{BjerrumBohr:2008ji}  and  discuss  their  consequences  for
unordered  amplitudes. In section~\ref{sec:generic} we confirm the
results obtained  using the world-line method  with on-shell unitarity
methods. Many advances in generalised unitarity techniques have been made
recently both in the context of analytic
computations~\cite{Britto:2004nc,Anastasiou:2006gt,Britto:2008vq,Forde:2007mi}
and for numerical
evaluations~\cite{Ossola:2006us,Giele:2008ve,Berger:2008sj}.
Using such analytic methods that exploit complex
analysis and factorisation properties~\cite{Forde:2007mi,Badger:2008cm} we
show  that the `no triangle'  property for  multi-photon amplitudes follows
from the behaviour of  the tree amplitudes as  the momentum flowing  in the cut
becomes large.  The appendices contains  the technical details  on the
evaluation of the cut amplitudes.

\section{One-loop amplitudes in the world-line approach}\label{sec:stringbased}
In this section we will describe the world-line approach of
ref.~\cite{BjerrumBohr:2008ji} for analysing the structure of
multi-leg one-loop amplitudes for colourless gauge theories. One
important property of colourless gauge theories is that the
tensorial structure associated with each different ordering of the
external legs is the \emph{same}. This leads to cancellations that
are not manifestly featured in the ordered amplitudes. Within the
traditional Feynman graph approach this fact is difficult to
implement but in the world-line approach it is particularly
transparent and leads to the specific set of reduction formul\ae\
derived in ref.~\cite{BjerrumBohr:2008ji}.

\subsection{One-loop amplitudes in colourless gauge theories}
The generic structure of colourless gauge theory amplitudes at
one-loop, {\it e.g.}, in QED or gravity can be given by the
following expression  based on a Schwinger proper-time
representation of the one-loop
amplitude~\cite{Bern:1991aq,Bern:1991an,Strassler:1992zr,Bern:1992ad,Schubert:2001he}
\begin{equation}
\label{e:Feyn} \cA^{\ol}_{n}=\int_0^\infty {dT\over T}\, T^{-D/2+n}
\int_0^1 d^{n-1}\nu\, \cP(h_i,k_i;\nu_i)\,
\exp(-T\,  Q_n)\,.
\end{equation}
In this equation $T$ is the one-loop proper-time and $\nu_i$ are Feynman
parameters associated with the external states of the amplitude.
These are integrated over the range $[0,1]$ with the following measure
of integration
\begin{equation}
\label{e:dnu} \int_0^1 d^{n-1}\nu\equiv  \prod_{i=1}^n \int_0^1
d\nu_i  \, {1\over n}\, \sum_{j=1}^n \delta(\nu_j=1)\,.
\end{equation}
We will use a symmetrised delta-function to fix the translational
invariance in the loop amplitude. The quantity $Q_n$ is defined by
\begin{equation}
\label{e:Qn}
Q_n\equiv \sum_{1\leq i<j \leq n} \, (k_i\cdot k_j)\,
G_B(\nu_i-\nu_j)\,. \end{equation}
The one-loop scalar world-line Green function $G_B(x)$ is defined by
\begin{equation}
G_B(x)= x^2-|x|\,,
\end{equation}
and is the solution to the one-dimensional Poisson equation
\begin{equation}
\partial_x^2 G_B(x) = \delta(x)\,.
\end{equation}
$G_B(x)$ is the infinite tension, $\alpha'\to0$, limit of the corresponding bosonic string
correlator (see~\cite{Bern:1991an,Strassler:1992zr,Bern:1992ad,Schubert:2001he}
for a justification of these rules)
\begin{equation}\label{e:XX}
G_B(\nu)=-{1\over6}+{1\over D}\lim_{\alpha'\to0}\,\eta_{mn} \big \langle x^m(\nu)
x^n(0)\big \rangle=\sum_{n\in\mathbb{Z}\backslash \{0\}} {1\over n^2}\,
e^{2i\pi n\, \nu}-{1\over6}\,.
\end{equation}
The constant $1/6$ in the above equation does not contribute to the
on-shell amplitudes because of momentum conservation.
See refs.~\cite{Schmidt:1994zj,Roland:1996im,Schubert:2001he,Dai:2006vj}
for a generalisation of the world-line formalism to higher-loop amplitudes.

We will also introduce the fermionic Green function
\begin{equation}\label{e:GF}
G_F(x)=\sg(x)\,.
\end{equation}
$G_F(x)$ is defined as the infinite tension, $\alpha'\to0$, limit of the world-line
correlator for fermions $\psi^m(\nu)$ with the anti-periodic boundary
conditions $\psi^m(\nu+1)=-\psi^m(\nu)$. The correlator $G_F(x)$ can be
expressed as
(see~\cite{Bern:1991an,Strassler:1992zr,Schubert:2001he} for a
justification of these rules)
\begin{equation}\label{e:ff}
G_F(\nu)={1\over D}\, \lim_{\alpha'\to0} \eta_{mn}
\big\langle \psi^m(\nu) \psi^n(0)\big\rangle_{A}= \,2\sum_{n\in\mathbb{Z}+\tfrac12}
{e^{2i\pi \, n\, \nu}\over n}\,.
\end{equation}

The  representation \eqref{e:Feyn} of  the one-loop  amplitude can  be  obtained by
considering  a Schwinger representation  of the  corresponding Feynman
integrals. For  instance by an exponentiation of the  propagators of the loop
amplitude  with   the  external   states   arranged   in  the   order
$\{1,2,\dots,n\}$ along the loop one can write
\begin{eqnarray}
  \prod_{i=1}^n {1\over (\ell-k_{1\cdots i})^2}&=&     \int_0^\infty    \prod_{i=1}^n     d\alpha_i    \,
  \exp\Big(-\sum_{i=1}^n \alpha_i \, (\ell-k_{1\cdots i})^2\Big)\\
\nn  &=&  \int_0^T  dT\,  T^{1-n}\,  \int_0^1
\prod_{i=1}^{n-1}da_i\, \exp\Big(-T \, (\ell - K_{[n]})^2-T\, Q_n\Big)\,,
\end{eqnarray}
where $k_{1\cdots i}=k_1+k_2+\cdots+k_i$ and the rescaled Schwinger parameters $a_i=\alpha_i/T$ are related to
the $\nu_i$ in eq.~\eqref{e:Feyn} by
\begin{equation}
  \nu_i=\sum_{j=1}^i a_j\ .
\end{equation}
As in~\cite{BjerrumBohr:2008ji} we use $\sigma$ to denote a given
ordering. ($\sigma$ is defined as a given permutation of the $n$
external legs $\{k_{\sigma(1)},\dots,k_{\sigma(n)}\}$). In this
notation the mapping between the $\nu_i$ and $a_i$ variables is
given by
\begin{equation}
\nu_i=\sum_{j=1}^i \, a_{\sigma(j)}\,.
\end{equation}
In the above representation one sees that the loop momentum is given
by the total inflow of external momenta
\begin{equation}
  \label{e:Kn}
  K_{[n]}= \sum_{j=1}^n \, k_j\,\nu_j\,.
\end{equation}
In  this  representation a  power  of  loop  momentum $\ell\cdot  k_i$
appearing in the numerator of the Feynman integral has the following
representation
\begin{eqnarray}\label{e:LtoK}
2 \, k_i\cdot K_{[n]} &=&2\,\sum_{j=1}^n (k_i\cdot k_j)\, \nu_j
 =-\partial_{\nu_i} Q_n+ 2\, \sum_{j=1}^n (k_i\cdot k_j) \,
G_F(\nu_i-\nu_j)\,.
\end{eqnarray}
This shows  that in  the world-line representation  the powers  of loop
momenta  in  the  amplitude   are  counted  by  the  first  derivative
$\partial_{\nu_i}Q_n$.
Following          the           strategy          defined          in
eq.~\cite{BjerrumBohr:2008vc,BjerrumBohr:2008ji} we
expand the polarisations of the external states in a basis of
independent momenta
\begin{equation}\label{e:cij}
h_i=\sum_{j=1}^{n-1} c_i{}^j \, k_j+ q^\perp\,.
\end{equation}
Here $q^\perp$ is a vector orthogonal to the $(n-1)$ linearly
independent external momenta. For an amplitude with $n>4$ external
legs the momentum $q^\perp$ is only needed in dimensions $D>4$. One
employs an identical definition for the $\bar h_i$ polarisations.
Using the relation~(\ref{e:Qn}) one easily derives that
\begin{eqnarray}\label{e:htoK}
2\,  h_i\cdot K_{[n]}&=&2\,  \sum_{r=1}^{n-1} \,  c_i^r\,
k_r\cdot K_{[n]}
 =\sum_{r=1}^{n-1}  c_i{}^r\, \Big[-\partial_{\nu_r}  Q_n+  \sum_{j=1}^n (k_r\cdot
k_j) \, G_F(\nu_r-\nu_j)\Big]\,.
\end{eqnarray}
It should be noticed as well that the second derivative on $Q_n$ is
given by
\begin{eqnarray}
\nn   \partial_{\nu_i} \partial_{\nu_j} Q_n&=  (k_i\cdot k_j)\,  \partial_{\nu_i}\partial_{\nu_j}
G_B(\nu_i-\nu_j)\qquad &\textrm{no~sum~over}\ i, j\,\\
&=  2(k_i\cdot k_j)\,  \big(\delta(\nu_i-\nu_j)-1\big)\qquad
&\textrm{no~sum~over}\ i, j\,,
\label{e:ddQ}\end{eqnarray}
and does not contain any powers of loop momenta.
The delta-function  $\delta(\nu_i-\nu_j)$ in the above expression
pinches  two  of the  external  legs.  The  constant arises  from  the
zero-mode contribution to the world-line Green function.

\medskip

The dependence on the external polarisations $h_i$ and momenta $k_i$
with $1\leq i\leq n$ is given by the function $\cP(h_i,k_i;\nu_i)$
which for the massless QED amplitude takes the form

\begin{equation}
\cP(h_i,k_i;\nu_i)=  \prod_{i=1}^n \int d\theta_i
\exp(\mathcal{F})\Big|_{\textrm{linear~in}~h_i}\,.
\end{equation}
Here  $\theta_i^\alpha$ are  $n$ Grassmann  variables that  we discuss
below, and one
has to keep only the terms linear in each of the polarisations
$h_i$  of the  external  states (with  $1\leq  i\leq  n$).
The  factor $\mathcal{F}$ is defined by
\begin{eqnarray}
\nn \mathcal{F}&=&
{1\over2\, T}\sum_{i\neq j} (h_{i}\cdot h_{j}) \,\theta_i\theta_j\, \partial_{\nu_i}\partial_{\nu_j}G_B(\nu_i-\nu_j)
+{i\over2}\sum_{i\neq    j}     \big(k_{i}\cdot
h_{j}\,\theta_i-k_{j}\cdot h_{i}\,\theta_j\big) \partial_{i}G_B(\nu_i-\nu_j)\\
\nn &+&
{1\over2}\sum_{i\neq j} (h_{i}\cdot h_{j}) \, G_{F}(\nu_{i}-\nu_j)
-{i\over2}\sum_{i\neq j} \big(k_{i}\cdot h_{j}\, \theta_{j}-k_{j}\cdot h_{i}\, \theta_{i}\big)\, G_{F}(\nu_i-\nu_j)\\
\label{e:cF} &+&{1\over2}\sum_{i\neq j}
\theta_{i}\theta_{j}\,(k_{i}\cdot k_{j})\, G_{F}(\nu_i-\nu_j)\,.
\end{eqnarray}
This factor is derived by  considering the correlation function of $n$ vertex
operators for a $U(1)$ gauge boson
\begin{equation}
\label{e:vop} V_i = (h_i\cdot\partial x +i k_i\cdot \psi \,
h_i\cdot\psi)\, e^{ik_i\cdot x}\,,
\end{equation}
as
\begin{equation}
\langle V_1 \cdots V_n\rangle = \exp({\mathcal{F}})\,
\exp(-T\, Q_n)\,.
\end{equation}
The expression~(\ref{e:cF}) has been written introducing the fermionic
variable $\theta_i$
\begin{eqnarray}\label{e:vop2}
V_i&=& \int d\theta_i\,h_i\cdot DX \, e^{ik_i\cdot X}=\int d\theta_i  \, \exp\left( h_i\cdot DX+ik_i\cdot x
\right)\Big|_{\textrm{linear~in}~h_i}\\
\nn&=&\int d\theta_i  \, \exp\big(\theta_i\,
(h_i\cdot\partial x)+h_i\cdot\psi+ \theta_i \,
(ik_i\cdot \psi)+ik_i\cdot x \big)\Big|_{\textrm{linear~in}~h_i}\,,
\end{eqnarray}
where $\theta$ is a fermionic variable in the $N=1$ world-line
formalism and $X$  is  a superfield  $X^m= x^m+\theta\psi^m$ with
the fermionic derivative $D  =
\partial_\theta - \theta\partial_\nu$
(see ref.~\cite{Brink:1976sc,Schubert:2001he,Dai:2006vj} for further
details).
The contractions between the world-line fields are done using the
correlators in eqs.~(\ref{e:XX}) and~(\ref{e:ff}).

\medskip

Using the relations in eqs.~\eqref{e:LtoK},~\eqref{e:cij}
and~\eqref{e:htoK} one can show~\cite{BjerrumBohr:2008ji} that the
generic form of a QED amplitude is given
 by  the  sum  of  unordered  $n$-point  integrals  $\cI_n[\I_r,\J_s]$
 evaluated in $D$ dimensions
where the positions of the external states are freely integrated
over the loop
\begin{equation}
\label{e:defcI} \cI_n[\I_r,\J_s]=    \int_0^1   d^{n-1}\nu    \,
Q_n^{D/2-n}   \,
\prod_{i\in\I_r} \partial_iQ_n\, \prod_{x\in \J_s}
G_F(x)\,,
\end{equation}
with gauge invariant tensorial coefficients $t^l_{r,s}$ built from the external momenta and polarisations
\begin{equation}
\label{e:QEDamp}
\cA^\ol_n=\sum_{u=0}^{n/2}\sum_{l=0}^u\sum_{r+s+2l=n} \, t^l_{r,s}
\,\cI^{[D+2(u-l)]}_{n-l}[\I_r,\J_s]\,.
\end{equation}
Here $\I_r=\{i_1,\dots,i_r\}$ is a set of $r$ indices of the
external states, and
$\J_s=\{\nu_{j_1}-\nu_{l_1},\dots,\nu_{j_s}-\nu_{l_s}\}$ is a set of
$s$ differences  of the positions  of the external states.  Because of
the   zero-mode  contributions   to  the   propagator   $G_B(x)$  (see
eq.~(\ref{e:ddQ})) the expression
involves
integrals that are evaluated in a dimension
different  from  $D$  that  we  denote by  introducing  a  superscript
indicating the dimension where the integral are evaluated,
e.g., $\cI_n^{[D+2u]}[\I_r,\J_s]$.

Using the relation in eq.~\eqref{e:LtoK} one can deduce that the
number of loop momenta in a one-loop $n$-photon amplitude in QED
(given by eq.~(\ref{e:QEDamp})) satisfies the constraint
\begin{equation}
\label{e:rsQED} r+s\leq n \ .
\end{equation}
We will also quickly review the result for the graviton amplitude as
given in~\cite{BjerrumBohr:2008ji}. The generic form of an
$n$-graviton amplitude in $\cN=8$ supergravity is given by
\begin{equation}
\label{e:SugraDef}
\cM_n^\ol =\sum_{u=0}^{n/2}\sum_{l=0}^u\,\sum_{r+s+2l=2n-8}\,
t^l_{r,s}\,\cI^{[D+2(u-l)]}_{n-l}[\I_r,\J_s]\,.
\end{equation}
This expression displays that the $n$-graviton one-loop amplitude
has at most $2n-8$ powers of the loop momentum and satisfies,
\begin{equation}
\label{e:rsSUGRA} r+s\leq 2n-8\,.
\end{equation}
The upper  bound arises  because the  two-derivative nature  of the
gravitational vertex  imply that the   amplitude has at  most $2n$
powers of loop momentum and eight powers of the loop momentum are cancelled by the integration over the sixteen
fermionic zero modes.

\medskip

Before we close this section we would like to make a few remarks:

\begin{itemize}

\item In computations of \emph{colourless}
one-loop amplitudes for gauge theories in the world-line approach
{\it all different orderings} of legs in eq.~(\ref{e:Feyn}) have the
\emph{same} tensorial structure. This particular point makes
colourless gauge theory amplitudes special and makes it
possible~\cite{BjerrumBohr:2008vc,BjerrumBohr:2008ji} to  reduce the
amplitudes to a form consisting of a sum of $\cI_n^{[D+2u]}[\I_r,\J_s]$
integrals as given in eqs.~\eqref{e:QEDamp} and~\eqref{e:SugraDef}.

\item Expressions for QED and supergravity amplitudes in
eqs.~(\ref{e:QEDamp}) and~(\ref{e:SugraDef}) contain non-analytic
functions featuring absolute numerical values of differences between
the $\nu$ variables as well as sign functions in the definition of
$G_F(x)$ in eq.~(\ref{e:GF}). These non-analytic functions are
lifted when the loop integral is formally evaluated and that splits
the integrals up into sums of different regions of analyticity of
the amplitude. The sum over these different regions of analyticity
is in direct correspondence with the sum over different physical
orderings of the amplitude.

\item The representation of the one-loop amplitude in massless QED and
supergravity given in~(\ref{e:Feyn}) is readily obtained by
considering the infinite tension limit $\alpha'\to0$ of the corresponding closed
one-loop amplitude.
No massive string modes play a role in these
computations~\cite{Bern:1991an,Strassler:1992zr,Bern:1992ad,Schubert:2001he}.
An extension of this world-line construction to
higher-loop amplitudes  in $\cN=8$
supergravity would give a field theoretic justification of the behaviour
of  the  multi-loop  four-graviton  amplitude  derived  using  string
theory~\cite{Berkovits:2006vc} and dualities
in~\cite{Green:2006gt,Green:2006yu,Green:2008bf}.
\end{itemize}

\subsection{The reduction formul\ae}\label{sec:RedFor}

The integrals forming the building blocks of the QED amplitude in
eq.~(\ref{e:QEDamp}) and the supergravity amplitudes in
eq.~(\ref{e:SugraDef}) satisfy  new types of reduction formul\ae\
that were derived in~\cite{BjerrumBohr:2008ji}. We will review these
in this section.

The basic building block of colourless gauge theories are the unordered
scalar $n$-point integrals
\begin{equation}\label{e:DefInnoJs}
\cI_n[\I_{r+1}]= \int_0^1     d^{n-1}\nu\, Q_n^{D/2-n} \prod_{i\in
\I_{r+1}}    \partial_iQ_n\,.
\end{equation}
Here $\I_{r+1}\equiv\{i_1,\dots,i_{r+1}\}$ is a set of $r+1$ distinct
indices taking values in $\{1,\dots,n\}$. It was shown
in~\cite{BjerrumBohr:2008ji} that, by integration by parts, these integrals
satisfy the reduction formul\ae\
\begin{eqnarray}
&& \cI_n[\I_{r+1}]=\nonumber\\
\label{e:Step1}
&& {2\over
D/2 - n + 1} \Bigg[               \sum_{
 j\in \I_{r-m+1}}
(k_{i_{r+1}}\cdot k_j)
  \left( -\cI_{n-1}^{(i_{r+1}j)}[\I^{(j)}_{r-1}]+\cI_n^{[D+2]}[\I^{(j)}_{r-1}]\right)\\
\nonumber &&+(m-1)\sum_{s=1}^n (k_{i_{r+1}}\cdot
k_s)\,\cI_{n-1}^{(i_{r+1}s)}[\I^{(r+1)}_{r-1}]\Bigg]\,.
\end{eqnarray}
We see that $\cI_n[\I_{r+1}]$ can be expressed as a sum of the
dimension shifted integrals $\cI^{[D+2]}_n[\I_{r-1}]$
and the one-mass $(n-1)$-point
integrals
\begin{equation}
\cI_{n-1}^{(ij)}[\I^{(j)}_{r-1}] \equiv  \int_0^1d^{n-1}\nu\,
Q_n^{D/2-n}\, \delta(\nu_i-\nu_j)  \prod_{s\in
\I^{(j)}_{r-1}}
\partial_{\nu_s}Q_n\,.
\end{equation}
Integrals with more than one mass are
defined in the same way with several delta function insertions.
The boundary term is vanishing because of the 1-periodicity of $Q_n$ in each of the
$\nu_i$ variables, $Q_n(\nu_1,\dots,\nu_i+1,\dots)=Q_n(\nu_1,
\dots,\nu_i,\dots)$ since $G_B(1-x)=G_B(x)$ for
$0\leq  x\leq 1$ and $G_B(0)=G_B(1)=0$.

As in~\cite{BjerrumBohr:2008ji} the rule eq.~(\ref{e:Step1}) can be summarised as
\begin{equation}
\label{e:Red1} \cI_n[(\partial Q_n)^r]\rightsquigarrow
\cI^\m_{n-1}[(\partial Q_n)^{r-2}]+ \cI_n^{[D+2]}[(\partial Q_n)^{r-2}]\,,
\end{equation}
where  $\cI^\m_{n}$ denotes  a massive  $n$-point integral.  Using the
relations~\eqref{e:LtoK}  and~\eqref{e:htoK} between the  loop momenta
and the derivative of $Q_n$, $\ell\sim \partial_\nu
Q_n$, this relation implies that
two powers of loop momenta $\ell$ are cancelled at each step
\begin{equation}
 \cI_n[\ell^r]\rightsquigarrow
\cI^\m_{n-1}[\ell^{r-2}]+ \cI_n^{[D+2]}[\ell^{r-2}]\,.
\end{equation}

When some factors of $G_F(x)$ are present in the integrand we have
to distinguish between the following cases
\begin{itemize}
\item[$\triangleright$] If all the $i\in \I_r$ are such that $\nu_i$ is not an
argument of $G_F(x)$  for any  $x\in \J_s$,  then the  same
manipulations leading to eq.~(\ref{e:Step1})
apply with no changes.
\item[$\triangleright$] If   $i_{r+1}$ has multiplicity one in
$\I_{r+1}=\I_r\cup \{i_{r+1}\}$ with $i_{r+1}\not\in \I_r$ and
$\J_1=\{\nu_{i_{r+1}}-\nu_j\}$ then
\end{itemize}
\begin{eqnarray}\label{eqnuGF}
\nn&&\cI_n[\I_{r+1},\J_1]=
{1\over
  D/2-n+1} \int_0^1  d^{n-1}\nu\,\partial_{\nu_{i_{r+1}}}Q_n^{D/2-n+1}
\, G_F(\nu_{i_{r+1}}-\nu_j)\,
\prod_{i\in\I_r}
\partial_iQ_n\,.\\
\end{eqnarray}
This leads, after integration by parts, to
\begin{eqnarray}
\cI_n[\I_{r+1},\J_1]&=&{
2\over  D/2-n+1}\times\\
\nonumber   &&\Bigg[   \sum_{j\in\I_r}
(k_{i_{r+1}}\cdot
  k_j) \nonumber
(-\cI_{n-1}^{(i_{r+1}j)}[\I_{r-1}^{(j)},\J_1]+\cI_n^{[D+2]}[\I_{r-1}^{(j)},\J_1])\\
\nonumber&&+ \Big((n-1)\cI_{n-1}^{(i_{r+1}
j)}[\I_{r}]-\sum_{l=1}^n \cI_n^{(i_{k+1}l)}[\I_{r}^{(r+1)}]\Big)\Bigg]\,.
\end{eqnarray}

This expression is easily generalised to other cases, with higher multiplicity
of $i_{k+1}$ and with additional $G_F$ contributions.

As in~\cite{BjerrumBohr:2008ji} this rule can be summarised by
\begin{equation}
\label{e:Red2} \cI_n[(\partial Q_n)^r,G_F]\ \rightsquigarrow\
\cI_{n-1}^\m[(\partial Q_n)^{r-1}]+ \cI_{n-1}^\m[(\partial Q_n)^{r-2},G_F]
+\cI_{n}^{[D+2]}[(\partial Q_n)^{r-2},G_F]\,.
\end{equation}
Using the relations~\eqref{e:LtoK} and~\eqref{e:htoK} between the loop
momentum $\ell$ and the first derivative of $Q_n$,
$\ell\sim \partial_\nu Q_n$, this relation can be rewritten as
\begin{equation}
 \cI_n[\ell^r,G_F]\ \rightsquigarrow\
\cI_{n-1}^\m[\ell^{r-1}]+ \cI_{n-1}^\m[\ell^{r-2},G_F]
+\cI_{n}^{[D+2]}[\ell^{r-2},G_F]\,.
\end{equation}

\subsection{Reduction of unordered one-loop amplitudes}
As was explained in ref.~\cite{BjerrumBohr:2008ji}
because one-loop amplitudes of $\cN=8$ supergravity
takes the symbolic form
\begin{equation}
  \label{e:cPsusy}
\mathcal{M}_{n;1}=  \sum_{r+s+u=2n-\cN\atop 0\leq u\leq
n}\sum_{l=0}^u\, t^l_{r,s}\,\cI^{[D+2(u-l)]}_{n-l}[\I_r,\J_s]\,,
\end{equation}
and due to the the constraint $r+s\leq 2n-8$, all amplitudes
can eventually be reduced to scalar box integral functions.
In this section we apply the reduction formul\ae\ of eqs.~(\ref{e:Red1})
and~(\ref{e:Red2}) to the QED amplitude~(\ref{e:QEDamp}) and repeat
our analysis in ref.~\cite{BjerrumBohr:2008ji} for the QED case.

The structure of the derivative structure of the cubic $q\bar q\gamma$
vertex implies that an $n$-photon one-loop  amplitude has at
most $n$ powers of  loop momenta.
The generic  form of  the QED amplitude  is given by
\begin{equation}
\label{e:QEDamp2}
\cA^\ol_n=\sum_{u=0}^{n/2}\sum_{l=0}^u\sum_{r+s+2l=n} \, t^l_{r,s}
\,\cI^{[D+2(u-l)]}_{n-l}[\I_r,\J_s]\,.
\end{equation}
In QED we have the constraint
\begin{equation}
\label{e:rsQED2} r+s\leq n \,,
\end{equation}
in the decomposition of the amplitude.
In fact because of Furry's theorem stated in eq.~(\ref{e:Symm}) the non-vanishing amplitudes
will have an even number of external photon states $n=2m$.
We will comment more on this
 in section~\ref{sec:oneloopQED}.
Applying   the reduction formul\ae\ of the  previous section to the
contribution with the  highest  power  of  loop  momenta
$\cI_{n}[I_{n}]$ will reduce it to $\cI_{m}[\emptyset]$ after
$n/2=m$ steps of reductions plus a contribution from  dimension
shifted integrals.  All other contributions with less powers  of
loop momenta, {\it i.e.}, $r+s\leq n$ with $s\neq0$, will reduce as
well to $\cI_{m}[\emptyset]$ plus the contribution from  dimension
shifted integrals. The dimension shifted integrals are of the type
\begin{equation}\label{e:DD}
 \cI^{[4+2p]}_{n+p}[\emptyset]=
\int_0^\infty {d^4\ell
    d^{2p}\ell_\perp\over                    (2\pi)^{4+2p}}\,
  \prod_{i=1}^{n+p}{1\over (\ell-k_1-\cdots- k_i)^2+\ell_\perp^2}\,,
\end{equation}
with $D=4$ and $1\leq p \leq n$.
These integrals do not carry any ultra-violet nor infra-red
divergences and have the special structure of the loop momentum
being integrated in $4+2p$ dimensions with $p\geq1$ but with only
four dimensional external momenta.  The dimension shifted
contributions are an artifact of the reduction formul\ae\ and they
do {\it not} contribute to the the total physical amplitudes as
shown in~\cite{BjerrumBohr:2008ji}.

From this analysis we can conclude that one-loop amplitudes with $n>4$
external    photons    do    not    contains   any    scalar    bubble
integrals. Amplitudes  with $n>6$ external photons do  not contain any
scalar triangle  integrals and will be completely  specified by scalar
box integrals.  We will  confirm these results directly via on-shell
unitarity methods in the following sections.

\section{Multi-photon amplitudes in QED with the unitarity method}\label{sec:generic}
In the  previous section we analysed  the structure of the
$n$-photon one-loop amplitude in massless QED from a string based
world line formalism. We showed that the amplitudes with $n>4$ external
photons do not  contains any scalar bubble integrals and
that amplitudes with $n>6$ external photons do not contain any
scalar triangle integrals and hence are completely specified by
scalar box integrals.

These results are in agreement with the explicit evaluation of the
one-loop four-photon  amplitude in~\cite{Mahlon:1993fe}  and  the
six-photon amplitude  computation
in~\cite{Binoth:2007ca,Bernicot:2007hs}.

In this section we will consider multi-photon one-loop amplitudes in QED
at the  field theory  level. The aim is to verify
the above string-based arguments for the structure of the loop
amplitude in QED. Looking at a simpler theory than gravity, QED, we
hope to shed light on how cancellations between various orderings
in unordered field theories can take place.

\subsection{The one-loop multi-photon amplitude \label{sec:oneloopQED}}
\begin{figure}[ht]\begin{center}
\includegraphics[width=5cm]{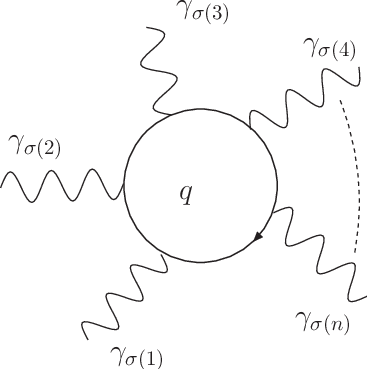}
\caption{\label{fig:loop} The one-loop $n$-photon amplitude in QED
is the sum over all
permutations of ordered photon lines attached to a fermion loop.}
\end{center}
\end{figure}

In this section we turn to the analysis of the photon
scattering at one-loop in massless QED
\begin{equation}
\gamma(k_1)+\cdots+\gamma(k_n)\to 0\,.
\end{equation}
All photon lines are attached to the massless fermion loop. As for
the case of the tree-level amplitude the one-loop amplitude can be
written as a sum over all the orderings of the external photon lines
\begin{equation}
A^\ol_{n;\Phi}(k_1,\dots,k_n)= \sum_{\sigma\in\mathfrak{S}_n} \,
\cA^\ol_{n;\Phi}(k_{\sigma(1)},\dots,k_{\sigma(n)})\,.
\end{equation}
Here $\cA^\ol_{n;\Phi}(k_{\sigma(1)},\dots,k_{\sigma(n)})$ is the ordered
one-loop amplitude  with $n$-photon  lines attached to a  fermion loop
for $\Phi=q$ or a complex scalar for $\Phi=\varphi$.

We will split the cut analysis of the photon amplitude into two parts,
one that can  be determined from standard four-dimensional  cuts and one involving
a $D$-dimensional component  that can be determined via massive  cuts. We
will write the full amplitude as:
\begin{equation}
    A^\ol_{n;q}(k_1,\ldots,k_n) = A^{\ol,CC}_{n;q}(k_1,\ldots,k_n)+R_n(k_1,\ldots,k_n)\,.
\end{equation}
Here the term $A^{\ol,CC}_n$ contains all divergences and logarithmic terms and the term $R_n$ contains all remaining
rational functions. The cut-constructible piece can be determined from cuts of photon
amplitudes with a massless internal fermion line.
By using the supersymmetric decomposition in terms of an $\mathcal{N}=1$ chiral
multiplet and a scalar loop contribution,
\begin{equation}
    A^\ol_{n;q}(k_1,\ldots,k_n) = A^{\ol}_{n;\cN=1}(k_1,\ldots,k_n) -
    A^{\ol}_{n;\varphi}(k_1,\ldots,k_n)\,,
\end{equation}
it is clear  that the rational terms originate from  the scalar loop $
A^{\ol}_{n;\varphi}(k_1,\ldots,k_n)$ since the
$\mathcal{N}=1$ amplitude is cut-constructible in four dimensions \cite{Bern:1994cg}. We can
therefore proceed to calculate these terms by computing cuts of the photon amplitude with a massive
scalar loop.\footnote{We note that it would also be possible to treat these contributions directly
by using a massive internal fermion loop within the $D$-dimensional cutting method as demonstrated in
reference \cite{Ellis:2008ir}. However, we choose to compute the rational terms from the massive scalar
loop since the cancellations from  the orderings of the external legs
 in the tree-level amplitudes can be made more explicit.}

Before  embarking on  a detailed  analysis  of the  structure of  the
multi-photon one-loop amplitudes in QED, we remark that because the
coupling of a photon to a pair of fermions or complex
scalar is odd under  charge conjugation, the ordered amplitudes satisfy
the following relation
\begin{equation}\label{e:Symm}
\cA_{n;\Phi}^\ol(k_1,\dots ,k_n)= (-1)^n \,
\cA^\ol_{n;\Phi}(k_n,\dots,k_1)\,.
\end{equation}
The total  amplitude summed over all orderings of the external
photon lines will therefore vanish for an odd number of external states. This is the
so-called Furry's theorem.  It is valid for fermionic  and scalar loop
and for 
$\cN=1$ super-QED amplitudes.

\subsection{The tree-level amplitudes}

Since the unitarity method constructs loop amplitudes
from products of on-shell tree-level
amplitudes, we will in this section first review the tree-level
amplitudes needed to derive the one-loop photon amplitude.

\subsubsection{The fermionic tree-level amplitudes}
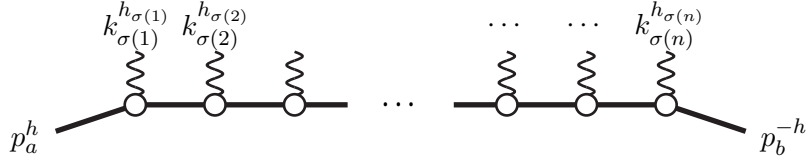
\begin{figure}[h]
\begin{center}
\begin{picture}(-100,-20)(50,80)
\Text(-142,18)[]{$p_a^h$} \Text(145,18)[]{$p_b^{-h}$}
\Text(0,30)[]{$\cdots$} \SetWidth{2} \Line(-130,20)(-100,30)
\Line(-100,30)(-70,30) \Line(-70,30)(-40,30)
\Line(-40,30)(-20,30) \SetWidth{1}
\Photon(-100,30)(-100,50){3}{3} \Photon(-70,30)(-70,50){3}{3}
\Photon(-40,30)(-40,50){3}{3}
\Text(-100,60)[]{$k_{\sigma(1)}^{h_{\sigma(1)}}$}
\Text(-70,60)[]{$k_{\sigma(2)}^{h_{\sigma(2)}}$}
\Text(100,60)[]{$\ k_{\sigma(n)}^{h_{\sigma(n)}}$}
\Text(40,60)[]{$\cdots$} \Text(70,60)[]{$\cdots$}
\Text(0,30)[]{$\cdots$} \SetWidth{2} \Line(130,20)(100,30)
\Line(100,30)(70,30) \Line(70,30)(40,30)
\Line(40,30)(20,30) \SetWidth{1} \Photon(100,30)(100,50){3}{3}
\Photon(70,30)(70,50){3}{3} \Photon(40,30)(40,50){3}{3} \SetWidth{1}
\CCirc(40,30){4}{Black}{White} \CCirc(70,30){4}{Black}{White}
\CCirc(-40,30){4}{Black}{White} \CCirc(-70,30){4}{Black}{White}
\CCirc(100,30){4}{Black}{White} \CCirc(-100,30){4}{Black}{White}
\end{picture}
\end{center}
\vspace{2cm} \caption{\label{fig:TreeFermion} Tree-level
$e^+e^-+n(\gamma)\to0$ Feynman diagram for the ordering
$\sigma\in\mathfrak{S}_n$. All the momenta are assumed to be
incoming. The helicity of the fermions is given by $h=\pm\frac12$.}
\end{figure}

The relevant  tree amplitudes  to the  contribution  of $A_{n;q}^{\ol,CC}$
coming from an internal massless fermion are those with photons coupling
to a pair of massless fermions. Since this
is an extremely simple process it is sufficient to use Feynman rules to
write down the amplitude as a sum over $n$ factorial permutations,
\begin{equation}
    \label{e:treeFeyn}
    A^\tree_{n;q}\big(p_a^h,p_b^{-h}; k_{1},\dots,k_{n}\big)=
    e^n\,\sum_{\sigma\in\mathfrak{S}_n} \,
    \cA^\tree_{n;q}\big(p_a^h,p_b^{-h};k_{\sigma(1)},\dots,k_{\sigma(n)}\big)\,.
\end{equation}
We refer to~\cite{Peskin:1995ev} for the QED Feynman rules.
All particles are considered to be incoming with momentum conservation
defined by $p_a+p_b+\sum_{i=1}^n k_i=0$.
We denote by the symbol $\mathfrak{S}_n$ the set of permutations of the $n$
objects and we denote a specific permutation by $\sigma\in\mathfrak{S}_n$. In
the above we have represented the momenta of the fermion and anti-fermion by
$p_a$ and $p_b$ respectively
while  the  $n$ photons  have been assigned  momenta  $k_i$ and  helicities
$h_i$. The ordered tree contribution, $\cA^\tree$, from an
individual Feynman diagram can in all generality be written as
\begin{equation}
\cA^\tree_{n;q}\big(p_a^h,p_b^{-h}; k_{\sigma(1)},\dots,k_{\sigma(n)}\big)=
\bar u_h(p_a) \sla \epsilon_{\sigma(1)}\,{i\sla q_1\over
q_{1}^2}\sla \epsilon_{\sigma(2)}{i \sla  q_2\over q_{2}^2} \cdots
\sla \epsilon_{\sigma(n-1)}{i\sla  q_{n-1}\over q_{n-1}^2}\sla
\epsilon_{\sigma(n)} u_{-h}(p_b)\,.
\end{equation}
Here $\sla v=\gamma^\mu v_\mu$ and  $u_{2h}(p)$ is the polarisation of
the fermion with helicity $h=\pm\frac12$.

The polarisation of the $i^{\rm th}$ photon is denoted by $\epsilon_i$, and
$q_i$ is the momentum flowing between leg $i$ and $i+1$
\begin{equation}
\label{e:qi} q_{i}= K_{\sigma(i)} +p_a, \qquad {\rm \ where \
}\qquad K_{\sigma(i)}\equiv \sum_{r=1}^i k_{\sigma(r)}\,.
\end{equation}
Using the conventions and notation introduced in appendix~\ref{sec:conventions}
one arrives at the following
Feynman representation of the tree-level amplitude~(\ref{e:treeFeyn})
given in~\cite{Kleiss:1986qc}
\begin{eqnarray}
\nn A_{n;q}^{\tree}\big(p_a^-,p_b^+;k_1,\ldots,k_n\big)&
=&{(-e/\sqrt2)^n\over \prod_{i=1}^n \langle p_{\rm
ref}^{i,-h_i}|k_i^{h_i}\rangle}\, \sum_{\sigma\in\mathfrak{S}_n}
\spa{a_{\sigma(1)}}.{p_a} \spb{p_b}.{b_{\sigma(n)}}
\prod_{i=1}^{n-1} {\spab{a_{\sigma(i+1)}}.{q_i}.{b_{\sigma(i)}}\over
q_i^2}\,.\\
\label{eq:ftree}
\end{eqnarray}
Here $\langle p^-|q^+\rangle=\langle pq \rangle$ and $\langle
p^+|q^-\rangle=[pq]$. In the above expression we have made the
helicity choice $h=\frac12$. The other amplitude with $h=-\frac12$
can be obtained by charge conjugation. The reference momentum $p^i_{\rm
ref}$ of the $i^{\rm th}$ photon is an arbitrary light-like momentum. It
cancels in the physical amplitude. We have employed the same
notation as introduced in~\cite{Kleiss:1986qc}
\begin{equation}
a_i={1+h_i\over2}\,p^i_{\rm    ref}+{1-h_i\over2}\,k_i,    \qquad
b_i={1+h_i\over2}\, k_i+{1-h_i\over2}\, p^i_{\rm ref}\,.
\end{equation}
For a generic photon helicity configuration this sum would contain
$n$ factorial terms. Choosing the reference momentum of the $h=+1$
helicity photons to be $p_a$  and the reference momentum of the
$h=-1$ helicity photons to be $p_b$ so that
\begin{equation}
\label{e:pref} p^i_{\rm ref}= {1+h_i\over 2}\, p_a+ {1-h_i\over 2}\,
p_b\,,
\end{equation}
one sees that the amplitude with $n_+$ photons of helicity $h_+=+1$ and
$n_-$ photons of helicity $h_-=-1$ has only  $n_+ \times n_-
\times (n_++n_--2)!$ non-zero contributions in the
sum~(\ref{e:treeFeyn}) or~(\ref{eq:ftree}).

With  the  choice of  reference  momentum eq.~(\ref{e:pref}) one sees that,
if all the helicities of the photons are $h=+1$ or all are of helicity
$h=-1$ each   term  in  the sum~(\ref{e:treeFeyn}) vanish. This is in
agreement with the supersymmetric Ward
identities~\cite{Grisaru:1976vm,Parke:1986gb,Berends:1987me,Mangano:1990by} for
supersymmetric QED.

Choosing the specific MHV helicity configuration (with one negative
helicity photon in the $1^{\rm st}$ position and all the rest positive), the
amplitude takes  the form of the  sum of permutations of ordered
MHV Parke-Taylor~\cite{Parke:1986gb} amplitudes

\begin{equation}
  \label{e:KSMHVbis}
 A^\tree_{n;q}\big(p_a^-,p_b^+;k_1^-,k_2^+,\dots,k_n^+\big)=
{e^n\over 2^{n\over2}}\,{\spa{p_a}.{k_i}^3\spa{p_b}.{k_i}\over
\spa{p_a}.{p_b}^{2}} \sum_{\sigma\in\mathfrak{S}_n} {1\over
  \spa{p_a}.{k_{\sigma(1)}}\spa{k_{\sigma(1)}}.{k_{\sigma(2)}}\cdots
  \spa{k_{\sigma(n)}}.{p_b}}\,.
\end{equation}
By making use of the eikonal identity given in~\cite{Berends:1987me}
\begin{equation}\label{e:Sumperm}
  \sum_{\sigma\in\mathfrak{S}_{n}} {\an[k_a,k_b]\over
    \an[k_a,k_{\sigma(1)}]\an[k_{\sigma(2)},k_{\sigma(3)}]\cdots
    \an[k_{\sigma(n)},k_b]}=  \prod_{1\leq  r\leq  n}
  {\an[k_a,k_b]\over \an[k_i,k_r]\an[k_j,k_r]}\,,
\end{equation}
one obtains the compact form
\begin{equation}
\label{e:KSMHV}
A^\tree_{n;q}\big(p_a^-,p_b^+;k_1^+,\dots,k_{i-1}^+,k_i^-,k_{i+1}^+,\dots,k_n^+\big)=
{e^n\over
2^{n\over2}}\,{\spa{p_a}.{p_b}^{n-2}\spa{p_a}.{k_i}^3\spa{p_b}.{k_i}\over
\prod_{j=1}^n \spa{p_a}.{k_j} \spa{p_b}.{k_j}}\,.
\end{equation}

It was shown in ref.~\cite{Ozeren:2005mp} that the $N^k$MHV amplitude can
be constructed using a CSW construction~\cite{Cachazo:2004kj} or via
BCFW~\cite{Britto:2004ap} recursion relations.

\subsubsection{Massive scalar tree amplitudes}\label{sec:massive}
\begin{figure}[h]
\begin{center}
\begin{picture}(-100,-20)(50,80)
\Text(-192,18)[]{$p_a^h$}

\Text(205,18)[]{$p_b^{-h}$} \Text(0,30)[]{$\cdots$} \SetWidth{2}
\DashLine(-180,20)(-150,30){3} \DashLine(-150,30)(-120,30){3}
\DashLine(-120,30)(-90,30){3} \DashLine(-90,30)(-70,30){3}
\SetWidth{1} \Photon(-150,30)(-150,50){3}{3}
\Photon(-120,30)(-120,50){3}{3} \Photon(-90,30)(-90,50){3}{3}
\Text(-150,60)[]{$k_{\sigma(1)}^{h_{\sigma(1)}}$}
\Text(-120,60)[]{$k_{\sigma(2)}^{h_{\sigma(2)}}$}
\Text(-90,60)[]{$k_{\sigma(3)}^{h_{\sigma(3)}}$}
\Text(160,60)[]{$k_{\sigma(n)}^{h_{\sigma(n)}}$}
\Text(105,30)[]{$\cdots$} \Text(130,60)[]{$\cdots$}
\Text(53,30)[]{$\cdots$}

\SetWidth{2} \Text(-55,30)[]{$\cdots$} \DashLine(-30,30)(-45,30){3}
\DashLine(-30,30)(-15,30){3} \SetWidth{1}
\Photon(-30,30)(-40,50){3}{3} \Photon(-30,30)(-20,50){3}{3}
\CCirc(-30,30){4}{Black}{White}
\Text(-45,60)[]{$k_{\sigma(i)}^{h_{\sigma(i)}}$}
\Text(-10,60)[]{$k_{\sigma(i+1)}^{h_{\sigma(i+1)}}$}

\SetWidth{2} \DashLine(25,30)(10,30){3} \DashLine(25,30)(40,30){3}
\SetWidth{1} \Photon(25,30)(25,50){3}{3}
\CCirc(25,30){4}{Black}{White} \Text(25,60)[]{$\cdots$}

\SetWidth{2}
\DashLine(78,30)(93,30){3} \DashLine(78,30)(63,30){3} \SetWidth{1}
\Photon(78,30)(88,50){3}{3} \Photon(78,30)(68,50){3}{3}
\Text(78,60)[]{$\cdots$} \CCirc(78,30){4}{Black}{White}

\SetWidth{2} \DashLine(190,20)(160,30){3}
\DashLine(130,30)(115,30){3} \DashLine(160,30)(130,30){3}
\SetWidth{1} \Photon(160,30)(160,50){3}{3}
\Photon(130,30)(130,50){3}{3}

\SetWidth{1} \CCirc(130,30){4}{Black}{White}
\CCirc(-90,30){4}{Black}{White} \CCirc(-120,30){4}{Black}{White}
\CCirc(160,30){4}{Black}{White} \CCirc(-150,30){4}{Black}{White}
\end{picture}
\end{center}
\vspace{2cm} \caption{\label{fig:TreeScalar} Tree-level scalar
Feynman diagram.}
\end{figure}
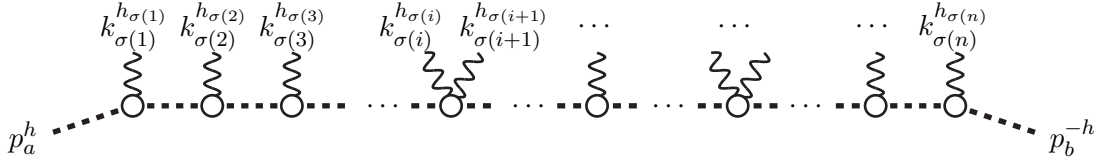

In this section we examine the massive scalar amplitudes
\begin{equation}
\varphi(p_a)+\varphi^*(p_b)+ \gamma(k_1)+\cdots +\gamma(k_n)\to 0\,.
\end{equation}
The tree-level amplitude with $n$ external photons attached to a
scalar line is built from cubic and quartic vertices of the
QED Lagrangian (again we will refer to ref.~\cite{Peskin:1995ev}
for details on the Feynman rules). The
amplitude is the sum over the permutations
\begin{equation}
A^\tree_{n;\varphi}\big(p_a^0,p_b^0;k_1,\dots,k_n\big)=e^n \,
\sum_{\sigma\in\mathfrak{S}_n}\,
\cA^\tree_{n;\varphi}\big(p_a^0,p_b^0;k_{\sigma(1)},\dots,k_{\sigma(n)}\big)\,,
\end{equation}
of an amplitude defined from the partition of the $n$ ordered external
legs partitioned in group of at most length two
\begin{equation}
\cA^\tree_{n;\varphi}\big(p_a^0,p_b^0;k_{\sigma(1)},\dots,k_{\sigma(n)}\big)
= \sum_{a_1+\cdots+ a_r=n\atop
   a_k\in\{1, 2\}}\,  \prod_{s=1}^r \, {\epsilon_{\sigma(a_1+\cdots
     + a_{s-1}+1)}\cdot H(a_s)\over
     (p_a+\sum_{j=1}^{a_1+\cdots+a_s}k_{\sigma(j)})^2-\mu^2}\,,
     \label{eq:Stree}
\end{equation}
with
\begin{equation}
  H(a_s)=
  \begin{cases}
    q+\sum_{j=1}^{a_1+\cdots   +a_{s-1}}  k_{\sigma(j)}&  \textrm{if}\
    a_s=1\\
\epsilon_{\sigma(a_1+\cdots   +a_{s})}&\textrm{if}\ a_s=2\,.
  \end{cases}\,
\end{equation}
Because  of the  cubic and  quartic vertices,  this amplitude  is
a much larger sum of terms than the fermionic case, since the scalar
$n$-photon tree-amplitude is a sum over $n!\times
F_{n+1}$ where $F_r$ is the Fibonacci number of order $r$ (such that
$F_0=F_1=1$ and $F_2=2$).\footnote{The tree-level
$A^\tree_{n;\varphi}$ photon scalar amplitude is constructed by
adding one external photon line  connected with a cubic vertex to
the $n-1$ amplitude  $A^\tree_{n-1;\varphi}$ or by  adding two
photon lines connected by the quartic vertex to the $n-2$ amplitude
$A^\tree_{n-2;\varphi}$.  Therefore the number  of ordered
amplitudes $F_n$ satisfies the Fibonacci recursion relation $F_n=F_{n-1}+F_{n-2}$.}

\subsection{BCFW shifts and large momentum scaling at tree-level \label{sec:BCFW}}

In this section we will analyse the large $z$ scaling
behaviour of photon tree-amplitudes under the BCFW shift~\cite{Britto:2004ap}. This is done for
cases of photons coupled to a massless fermion pair and a massive scalar pair which is useful for our
investigation of which scalar integrals appear in the $n$-photon
one-loop amplitude as described earlier in this section.

Cancellations in one-loop graviton scattering has already been
studied through relating the coefficients of a scalar integral basis
of the one-loop amplitude and the scaling behaviour of their
corresponding unitarity cuts, {\it i.e.}, products of on-shell tree
amplitudes, under BCFW
shifts~\cite{BjerrumBohr:2006yw,Bern:2007xj,ArkaniHamed:2008gz}.

For example the large $z$ limit of the BCFW shift of the cut propagator momenta
is related~\cite{Bern:2007xj} to the large $t$ limit in the triple cut for
triangles (see section~\ref{sec:triangle}) using Forde's parametrisation
of the cut loop momentum~\cite{Forde:2007mi}. Similarly
(see section~\ref{sec:bubble}) the large $z$ scaling is
related~\cite{Bern:2007xj} to the large $y^2/t$
scaling in the double cut in Forde's  parametrisation
of  the cut loop  momentum.
The large $z$ limit of the BCFW scaling behaviour for the cut tree amplitude
can be used directly to test if the one-loop amplitude
has any scalar triangle and bubble integrals.

Rational pieces in the amplitude can be probed for in a similar way
using $D$-dimensional unitarity
techniques~\cite{Bern:1995db,Bern:1996ja,Anastasiou:2006gt,Britto:2008vq,Giele:2008ve}.
One can then relate the large $z$ limit of the BCFW
shift to the large momentum limit of a massive
cut loop momentum following the methods of~\cite{Forde:2007mi,Badger:2008cm}.

\subsubsection{Large-$z$ scaling for the massless tree amplitudes}
Since the fermion line carries the loop momentum in the $n$-photon
amplitude the relevant BCFW shift for the fermion tree amplitudes
is that of shifting the quark, $p_a$, and anti-quark, $p_b$:
\begin{eqnarray}
\label{e:S1} \textrm{type}~s=+1, \qquad |\widehat p_a\rangle &\equiv
|p_a\rangle + z|p_b\rangle,\qquad
|\widehat p_b] &\equiv |p_b] - z|p_a]\,,\\
\label{e:S2} \textrm{type}~s=-1, \qquad|\widehat p_b\rangle &\equiv
|p_b\rangle + z|p_a\rangle,\qquad |\widehat p_a]&\equiv |p_a] -
z|p_b]\,.
\end{eqnarray}
Under these shifts the polarisation of the external fermions behave as
\begin{eqnarray}
\label{e:ushift} u_\pm(\widehat p_a) =u_\pm(p_a) + z\, {s\pm
1\over2} \, u_\pm(p_b), \qquad \nn u_\pm(\widehat p_b) =u_\pm(p_b) +
z\, {s\mp 1\over2}\, u_\pm(p_a)\,,
\end{eqnarray}
while the propagator factors
$\sla q_i= \sla K_{\sigma(i)}  + \sla p_a$ shift according to
\begin{equation}
\label{e:qshift}
 \widehat {\sla q}_i= \sla q_i+z \, \sla \pi\,,
\end{equation}
where $\pi$ is a light-like vector defined by
\begin{equation}
\label{e:pi} \sla \pi \equiv  {1+s\over2} \,\Big( |  p_b\rangle
[p_a| +  | p_a] \langle p_b| \Big) +{s-1\over2}\, \Big( |
p_a\rangle [p_b| + | p_b] \langle p_a|\Big)\,.
\end{equation}

The fermion tree amplitudes in eq.~\eqref{eq:ftree} and the scalar tree
amplitude of eq.~\eqref{eq:Stree} have the following behaviour in the large $z$ limit
\begin{equation}
\label{e:largez}
\lim_{z\to\infty} A^\tree_{n;q}\big(p_a^h,p_b^{-h};k_1,\dots,k_n\big)\sim
C^\infty_q\big(p_a^h,p_b^{-h};k_1,\dots,k_n\big) \times {z^{2h\, s}\over
z^{n-2}}\,,
\end{equation}
where   $s=+1$  for   the  shift~(\ref{e:S1}) and  $s=-1$   for  the
shift~(\ref{e:S2}) and $h=\pm\frac12$ for a fermion and $h=0$ for a
scalar.

We  have   checked  this   behaviour  numerically  for   all  helicity
configurations  up  to  $n=10$  photon  lines. For  the  case  of  the
fermionic tree amplitude a formal  proof of this behaviour is given in
the  appendix~\ref{sec:proof}.  The   case  for  the  massless  scalar
amplitude  follows directly  from  the fermionic  case  by using  the
$\cN=1$ super-QED Ward identities.

\subsubsection{Large-$z$ scaling of the massive scalar tree amplitudes}
To analyse the rational contributions to the  one-loop photon amplitudes we
must consider a BCFW shift of
the two massive scalar particles in the tree amplitude given by eq.~\eqref{eq:Stree}.

In order to solve the on-shell conditions for a recursive
construction of an amplitude while shifting two massive particles it
is necessary to define two additional massless vectors
\cite{Schwinn:2007ee}. We therefore define a pair of ``flattened''
vectors from a pair a massive vectors $p_a,p_b$ each with mass
$\mu$:
\begin{eqnarray}
  p_a= p_a^\flat+{\mu^2\over\gamma} p_b^\flat;\qquad
  p_b= p_b^\flat+{\mu^2\over\gamma} p_a^\flat\,,
\end{eqnarray}
where $(p_a^\flat)^2=0$ and $(p_b^\flat)^2=0$ and
\begin{equation}
\gamma=2(p_a^\flat\cdot p_b^\flat)=(p_a\cdot p_b)\,\left(1+ \sqrt{1-{\mu^4\over (p_a\cdot p_b)^2}}\right)\,,
\end{equation}
so    that   $p_a^\flat\to   p_a$    and   $p_b^\flat\to    p_b$   for
$\mu^2\to0$.
For fixed $\mu^2$ we define  the shift  as
\begin{eqnarray}
  |\widehat p_a^\flat\rangle &=&|p_a^\flat\rangle +z\, |p_b^\flat\rangle;\qquad
  |\widehat p_b^\flat] =|p_b^\flat] -z\, |p_a^\flat]\,,
\end{eqnarray}
implying that the original momenta are shifted according to
\begin{equation}\label{e:massshift}
\widehat{\sla    p}_a=    \sla    p_a   +z\,   \left(1-{\mu^2\over
\gamma}\right)\, |p_b^\flat\rangle[p_a^\flat|;\qquad
\widehat{\sla    p}_b=    \sla    p_b   -z\,   \left(1-{\mu^2\over
\gamma}\right)\, |p_a^\flat\rangle[p_b^\flat|\,.
\end{equation}
In  this  case  the  large   $z$  limit  of  the  shifted  propagators
in~(\ref{eq:Stree}) become
\begin{equation}
\lim_{z\to\infty}  {1\over (\widehat p_a+  K)^2-\mu^2} \sim  {1\over z
  \,(\pi\cdot K)} \, {1\over 1-\mu^2/\gamma}\,,
\end{equation}
where $\pi$  is defined  as in eq.~(\ref{e:pi})  with $p_a$  and $p_b$
replaced by $p_a^\flat$ and $p_b^\flat$ respectively. We have in this argument
used that
$ \pi\cdot  p_a= 0$.  At the leading  order in $\mu^2$  the asymptotic
value of the propagators take a similar form to the one appearing in the
massless case.  This indicates that the leading large $z$ behaviour of the
massive scalar tree amplitudes is the same as in the massless case
\begin{equation}
\lim_{z\to\infty}  A^\tree_{n;\varphi}\big(p_a^h,p_b^{-h};k_1,\dots, k_n\big)=
 {1\over
   z^{n-2}}\,C^\infty_\varphi\big(p_a^h,p_b^{-h};k_1,\dots,k_n|\mu^2\big)\,.
\label{e:largezmassive}
\end{equation}
We have checked this behaviour numerically up to  $n=8$ external
photons and for all helicity configurations.

The  coefficient of  $C^\infty_\varphi$ depends  on the  mass
$\mu^2$, and in the massless limit $\mu^2\to0$  the behaviour of a
massless scalar tree amplitude is recovered. In the particular cases
of the all-plus or all-minus photon helicity configuration, which
vanish     in    the     massless     limit, the contribution
$C^\infty_\varphi\big(p_a^h,p_b^{-h};k_1^+,\dots,k_n^+|\mu^2\big)=O(\mu^2)$.
It can be seen that these sub-leading  contributions  play  an
important  role  in  the analysis of the potential rational pieces
in appendix~\ref{sec:rationalterms}.

For large  $\mu$ we express the  massive momenta in  terms of massless
momenta using
\begin{equation}
  p_a=  \mu\,   p^\flat_a  +  {\mu\over\gamma}\,p^\flat_b;\qquad  p_b=
  {1\over \mu} \,p^\flat_b+ {\mu^3\over\gamma}\,p^\flat_a\,,
\end{equation}
with the same definition for $\gamma$. To write this solution we used
the freedom to  rescale the massless momenta $(p^\flat_a,p^\flat_b)\to
(\lambda\, p^\flat_a,\lambda^{-1}\,p^\flat_b)$. We have chosen a linear scaling
in   $\mu$   since   this   is    what   will   be   needed   in   the
appendix~\ref{sec:rationalterms}  for  analysing  the eventual rational  pieces
from boxes.

 For large $\mu$ we have that
\begin{equation}
  \lim_{\mu\to\infty} {\gamma\over\mu^2}= i\, \sg(p_a\cdot p_b)\,,
\end{equation}
and in this limit
\begin{equation}\label{e:largepa}
  p_a\sim \mu \, p^\flat_a; \qquad p_b\sim -i\mu\, {p_a\over\sg(p_a\cdot p_b)}\,.
\end{equation}
In this case the asymptotic form of the propagators is given by
\begin{equation}
  \lim_{\mu^2\to\infty}   {1\over  (p_a+K)^2-\mu^2}\sim   {1\over  2\,
    \mu\,(p_a^\flat \cdot K)}\,,
\end{equation}
which  is of  the same  form as  for the  BCFW shift  of  the massless
amplitude with $z\sim \mu$. In  this case the tree amplitudes have the
asymptotic behaviour given in eq.~(\ref{e:largezmassive}) with $z=\mu$.

\bigskip
These results are used in the appendix~\ref{sec:rationalterms} where the
rational piece contributions to the one-loop amplitude are analysed.

\subsubsection{Origin of the improved BCFW scaling behaviour \label{sec:sumdisc}}

In this section we examine the analytic structure of the
cancellations that give rise to the improved BCFW scaling behaviour
observed in the preceding sections. We make use of a specific gauge
choice which makes the cancellations manifest in each of the
contributing dia\-grams. However, we will see that this is not
sufficient in order to remove the role of the summation of external
orderings for anything but the simplest Abelian amplitudes.

The analysis in the previous sections (and the
appendix~\ref{sec:proof}) showed that the tree-amplitudes in QED are
extremely well behaved in the large $z$ limit of the BCFW shifts
\begin{equation}\label{e:largez2}
    \lim_{z\to\infty}\sum_{\sigma\in\mathfrak{S}_n}     \cA_n(z)    \sim
    {z^{2hs}\over z^{n-2}}\,.
\end{equation}
For the case of the massless fermion tree-amplitudes this property
can be proven diagram by diagram by using a special gauge choice.
However this technique is not sufficient to find the observed
scaling property in more complicated amplitudes containing higher
point interactions such as the massive scalar tree amplitude and
those in gravity. In the discussion below we give explicit examples
of how the sum over external orderings is responsible for the
remaining cancellations.

In a Feynman graph based analysis of the scalar tree-amplitudes one can observe
that the cubic vertex at worst scales like $O(z)$, the quartic vertex like $O(1)$ and that the
ordered amplitudes in eq.~(\ref{eq:Stree}) scale at worst like
\begin{equation}
\lim_{z\to\infty}\mathcal{A}^\tree_{n;\varphi}(z)\sim   z\,.
\end{equation}
By taking the transverse gauge $\pi\cdot \epsilon^{h_i}=0$ for all the
photons by setting the reference momenta to be $p_{\rm ref}=\pi$ where
$\pi$ is the light-like momentum defined in eq.~(\ref{e:pi}), then all
cubic vertices  are independent  of $z$ and  scale as $O(1)$.  In this
gauge the ordered amplitudes  have the large $z$ scaling $\cA_n(z)\sim
1/z^{n/2-1}$ dominated  by graphs with  the maximum number  of quartic
vertices.  The sum  over the orderings of external  legs improves this
behaviour to the optimal one~(\ref{e:largez2}) as can be seen from the
following four-photon  and five-photon  examples.  With the  choice of
gauge $p_{\rm ref}=\pi$ only the contractions between polarisations of
opposite helicities are non-vanishing
 \begin{equation}
   \epsilon^\pm_i(k_i,\pi)   \cdot  \epsilon^\pm_j(k_j,\pi)=0;  \qquad
   \epsilon^\pm_i(k_i,\pi) \cdot \epsilon^\mp_j(k_j,\pi)\neq 0\,.
 \end{equation}

\begin{figure}[ht]
  \centering
\includegraphics[width=15cm]{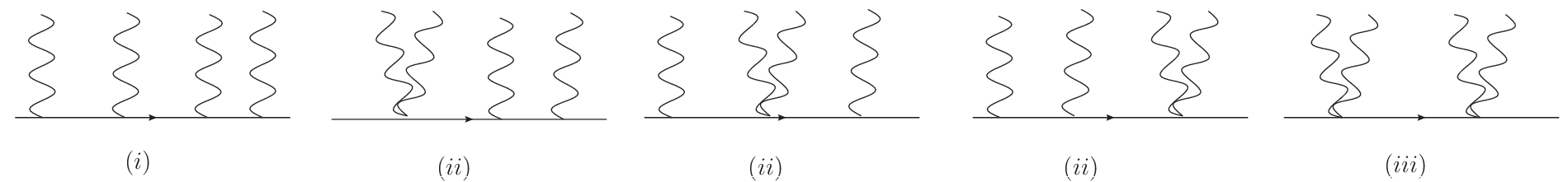}
  \caption{The  $e^-e^+\to 4\, \gamma$  tree-amplitude is  composed by
    the Feynman diagrams built from (i) four
three-point  vertices,
(ii)  two
three-point vertices and one four-point vertex, and
(iii) two four-point  vertices. The label of the  external photon have
to be symmetrically distributed over the photon lines.}
\label{fig:4ptscalar}
\end{figure}

We will be considering as a first example the four-photon amplitude with the helicity configuration
$(--++)$. For  a colourless  theory the orderings  of the  external legs
does  not  matter.  The  tree-amplitude is built from permutations of the four
topologies shown in figure~\ref{fig:4ptscalar}: (i) four
three-point  vertices, which  have the  large $z$  scaling $O(1/z^3)$,
(ii)  two
three-point  vertices  and  one  four-point  vertex,  which  scales  as
$O(1/z^2)$,
(iii) two four-point vertices, which has the scaling  $O(1/z)$.

Each ordered  contribution with two  four-point vertices has  a large
$z$ behaviour that is worse than the observed behaviour for the
total amplitude.
We  show that  this  bad scaling  is  cancelled in  the  sum over orderings
of external legs.

The contribution  from all the  four-point vertices to  the four-photon
amplitude is given by

\vskip -.6cm

\begin{align}
    \sum_{\sigma_2\times\sigma_2} &\frac{
    \varepsilon^{h_{\sigma(1)}}_{\sigma(1)}\cdot\varepsilon^{h_{\sigma(2)}}_{\sigma(2)}
    \varepsilon^{h_{\sigma(3)}}_{\sigma(3)}\cdot\varepsilon^{h_{\sigma(4)}}_{\sigma(4)}
    }{(\wh p_a+K_{\sigma(1)\sigma(2)})^2-\mu^2}\nonumber\\
    &= \frac{\varepsilon^-_1\cdot\varepsilon^+_3
    \varepsilon^-_2\cdot\varepsilon^+_4}{z\alpha\, \pi\cdot K_{13}}
    + \frac{\varepsilon^-_2\cdot\varepsilon^+_4
    \varepsilon^-_1\cdot\varepsilon^+_3}{z\alpha\,\pi\cdot K_{24}}
    + \frac{\varepsilon^-_1\cdot\varepsilon^+_4
    \varepsilon^-_2\cdot\varepsilon^+_3}{z\alpha\,\pi\cdot K_{14}}
    + \frac{\varepsilon^-_2\cdot\varepsilon^+_3
    \varepsilon^-_1\cdot\varepsilon^+_4}{z\alpha\,\pi\cdot K_{23}}
    + \mathcal{O}(z^{-2})\nonumber\\
 &= \varepsilon^-_1\cdot\varepsilon^+_3\,\frac{\pi\cdot K_{1234}}{z^2\alpha^2\,
   \pi\cdot K_{13}\,\pi\cdot K_{24}}
    + \varepsilon^-_1\cdot\varepsilon^+_4
\frac{\pi\cdot K_{1234}}{z^2\alpha^2\,\pi\cdot K_{14}\,\pi\cdot K_{23}}
    + \mathcal{O}(z^{-2})\nonumber\\
    &= \mathcal{O}(z^{-2})\,,
\end{align}
where $\sigma_2\times\sigma_2$  means that one has to sum  over the 2-cycle
decompositions of the permutations.
The cancellation  arises because of  momentum conservation $k_1+\cdots
+k_4=-p_a-p_b$ and via $\pi\cdot p_a=0$ and $\pi\cdot p_b=0$.

Using a more compact notation we have showed that
\begin{equation}
    (1^-,3^+)\cdot(2^-,4^+)+(2^-,4^+)\cdot(1^-,3^+)
    \underset{z\to\infty}{\to}\mathcal{O}(z^{-2})\,.
\end{equation}
In  the   case  of  the   five-photon  amplitude  with   the  helicity
configuration $(--+++)$, we find that the cancellations now involves all twelve different orderings:
\begin{align}
    &(3^+)\cdot(1^-,4^+)\cdot(2^-,5^+)
    +(3^+)\cdot(2^-,5^+)\cdot(1^-,4^+)\nonumber\\
    +&(1^-,4^+)\cdot(3^+)\cdot(2^-,5^+)
    +(2^-,5^+)\cdot(3^+)\cdot(1^-,4^+)\nonumber\\
    +&(1^-,4^+)\cdot(2^-,5^+)\cdot(3^+)
    +(2^-,5^+)\cdot(1^-,4^+)\cdot(3^+)+(4\leftrightarrow5)\nonumber\\
    \underset{z\to\infty}{\ \ \ \propto}&\ \frac{\varepsilon_3^+\cdot( 2p_a + K_{1245})}{z^2}\bigg(
    \frac{1}{\pi\cdot K_{14}\pi\cdot K_{25}}
    +\frac{1}{\pi\cdot K_{3}\pi\cdot K_{14}}
    +\frac{1}{\pi\cdot K_{3}\pi\cdot K_{25}}
    \bigg) + (4\leftrightarrow5)\nonumber\\
    =&\ \frac{\varepsilon_3^+\cdot( 2p_a + K_{1245})}{z^2}\,
    \frac{\pi\cdot K_{12345}}{\pi\cdot K_{14}\,\pi\cdot K_{3}\,\pi\cdot K_{25}}
   + (4\leftrightarrow5)
    \nonumber\\
    =&\ \mathcal{O}(z^{-3}).
\end{align}
We have used  momentum conservation  $k_1+\cdots  +k_5=-p_a-p_b$ and
that $\pi\cdot  p_a=0$ and $\pi\cdot p_b=0$.  The  first step requires
that the terms from the ($4\leftrightarrow 5$) exchange but the relation is
independent  of  the  momentum  appearing in  the  single  three-point
interaction. It  is therefore sufficient to show  that the five-photon
amplitudes scale as $1/z^3$ as required.

In  the  gravity  case  we  have a  similar  phenomenon.   The
multi-graviton vertices  na{\"\i}vely scale  as $z^2$ in the  large
$z$ limit  and the  ordered  Feynman  graphs scale  at  worst  like
$z^{n-1}\times  z^{f(h_1,h_2,s)}$. Here  $  f(h_1,h_2,s)\in[-4, 4]$
is an integer valued function of the polarisations  $h_{1,2}$ of the
shifted  legs and the type  of shift $s=\pm1$. By considering the
transverse gauge
\begin{equation}
 \pi^\mu \epsilon^{h_i}_{\mu\nu}(k_i,\pi) =0; \qquad
  \epsilon^{h_i}_{\mu\nu}(k_i,\pi) \, \pi^\nu = 0\,,
\end{equation}
and by setting  the reference momentum
of  the  unshifted  legs  to be $p_{\rm  ref}=\pi$,  the  multi-graviton
vertices that do not involve the two shifted legs scale
 at most as $O(z)$ and the worst scaling of the ordered Feynman graphs
is thus given by the $z$ dependence of the polarisation of the
shifted legs $z^2\,z^{f(h_1,h_2,s)}$.  The  sum   over  the
orderings of the external legs improves  the scaling behaviour of
the  total amplitude  to either  $z^{-2}$  or $z^6$  depending on  the
polarisation of the
shifted states~\cite{BjerrumBohr:2005jr,BjerrumBohr:2006yw,Benincasa:2007qj,ArkaniHamed:2008yf}.

\subsection{One-loop structure from large $z$ momentum scaling}

\begin{figure}[ht]
\centering
\includegraphics[width=10cm]{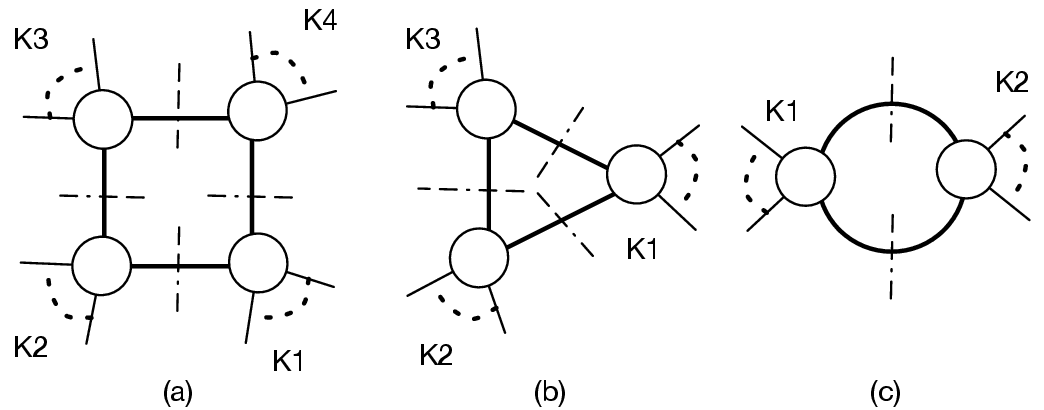}
\caption{ \label{fig:cut} Representation of (a) the quadruple cut
contributing to the coefficient $c_{4;K_1|K_2|K_3|K_4}$, (b) the
triple cut  contributing to the coefficient $c_{4;K_1|K_2|K_3}$ and
(c) the double cut contributing to the coefficient $c_{2;K_1|K_2}$
of the one-loop amplitude.}
\end{figure}

The generic decomposition of a $n$-photon one-loop amplitude in
dimensional regularisation with four-dimensional external momenta
is given by
\begin{eqnarray}
A^\ol_{n;q}&=& \sum_{\{K_i\}} c_{4;K_1|K_2|K_3|K_4} \, I^{(K_1|K_2|K_3|K_4)}_4
+ \sum_{\{K_i\}}
c_{3;K_1|K_2|K_3} \, I^{(K_1|K_2|K_3)}_3\\
\nn& +& \sum_{\{K_i\}} c_{2;K_1|K_2} \,
I^{(K_1|K_2)}_2 + R_n+\mathcal{O}(\epsilon)\,.
\end{eqnarray}
Here  $I^{(K_1|K_2|K_3|K_4)}_4$,  $I^{(K_1|K_2|K_3)}_3$,
$I^{(K_1|K_2)}_2$ and $R_n$ are scalar box, triangle, bubble
integrals and rational terms respectively evaluated in
$D=4-2\epsilon$  dimensions~\cite{Bern:1993kr,Ellis:2007qk}. The
coefficients $c^i_{4;K_1|K_2|K_3|K_4}$, $c^j_{3;K_1|K_2|K_3}$ and
$c^k_{2;K_1|K_2}$ are associated with the corresponding  scalar
integral  functions represented  in fig.~\ref{fig:cut}  where  $K_i$
are  the  sum of  the momenta at each vertex of the scalar integral
functions.

In this section we will outline the consequences of the large momentum
scaling at tree-level for the structure of the one-loop amplitude.
The details of our analysis can be found in appendix~\ref{app:cutanalysis}.

\subsubsection{Vanishing of triangle coefficients}

Parametrising the unfixed integration of the triple cut by a
complex parameter $t$ leaves the
triangle coefficients completely determined~\cite{Forde:2007mi}.
In the appendix~\ref{sec:triangle} we relate the large $t$-scaling of the
tree amplitudes in the cut to the large $z$ behaviour under the
BCFW shift (see as well ref.~\cite{Bern:2007xj}).  We show in the appendix
that the large $t$
scaling is  independent of  the helicity of  the state running  in the
loop ($h=\pm1/2$ for the fermion or $h=0$ for the scalar)
\begin{equation}
A_n^\tree \underset{t\to\infty}{\to} \frac{C_q^\infty(K_1,K_2)}{t^{n-2}}\,.
\end{equation}
We will show in the appendix~\ref{sec:triangle} that the scaling behaviour
of  the tree  amplitude in  the cut  leads us  directly to  the result
expected  from  the  world-line   analysis:  {\it namely that  all  triangle
  coefficients vanish  for one-loop $n>6$  amplitudes.}  This property
is unexpected  from na{\"\i}ve power counting  and from the
conjecture for the NNMHV amplitudes  of ref.~\cite{Bernicot:2007hs}.

\subsubsection{Vanishing of bubble coefficients}
As described in appendix~\ref{sec:bubble} the above scaling behaviour
gives sufficient information for concluding: {\it that all bubble
coefficients vanish for $n\geq 6$.} For the case of $n=5$
the coefficients vanish after the sum over the orderings of the external legs is performed.
This is in agreement with Furry's theorem.
The result that no scalar
bubble integral functions are present in
multi-photon one-loop amplitude with $n\geq 6$ legs
is in complete agreement with the world-line
formalism  of   section~\ref{sec:stringbased}.  Our  results  although
unexpected from na{\"\i}ve power counting also fits the analysis of the
MHV amplitude given in~\cite{Mahlon:1994dc}
and the result for the six-point NMHV amplitudes
computed in~\cite{Binoth:2007ca,Bernicot:2007hs}.

\subsubsection{Vanishing of rational terms}
In section~\ref{sec:rationalterms} we show that the $n$-photon
amplitude with $n\geq 5$ cannot have rational polynomial contributions.
The result of this analysis confirms the world-line string based result
{\it that there are no rational contributions for photon
amplitudes with $n>4$}.
From a field theory perspective, the rational polynomial contributions can be ruled out via
an analysis of the massive scalar tree amplitudes in the
large mass limit as described in ref.~\cite{Badger:2008cm}.

This is in perfect agreement with the direct
computation of Mahlon~\cite{Mahlon:1993fe} of the finite
helicity configurations:
\begin{eqnarray}
\nn A^\ol_{4;q}(k_1^+,k_2^+,k_3^+,k_4^+)&= &
e^4\,
{\sq[k_1,k_2]\sq[k_3,k_4]\over \an[k_1,k_2]\an[k_3,k_4]}\,,\\
\nn A^\ol_{4;q}(k_1^+,k_2^+,k_3^+,k_4^-)&= & e^4\,
{\sq[k_1,k_2]\sq[k_2,k_3]\an[k_3,k_1]\over \an[k_1,k_2]\an[k_2,k_3]\sq[k_3,k_1]}\,,\\
\nn
A^\ol_{n;q}(k_1^+,\dots,k_n^+)&=&0 \qquad \textrm{for}\ n\geq5\,,\\
 A^\ol_{n;q}(k_1^+,\dots,k_{n-1}^+,k_n^-)&=&0 \qquad \textrm{for}\ n\geq5\,.
\end{eqnarray}
It is also in agreement with the $n$-point MHV computation of
Bernicot et al.~\cite{Bernicot:2007hs} and the known
six-point computations of refs.~\cite{Binoth:2007ca,Bernicot:2007hs}.

\subsection{No-triangle property of one-loop photon amplitudes}
The vanishing of the coefficients described above leads us to the
expected result previously obtained via the
world-line analysis given by eq.~\eqref{e:QEDamp}.
The multi-photon fermionic or scalar one-loop amplitudes with $n\geq 8$
external photons satisfy a no triangle property, and hence contain
solely scalar box integral functions in $D=4-2\epsilon$
dimensions. The $n$ point amplitude can thus be written as
\begin{equation}\label{e:Ampn8}
A^\ol_{n;\Phi} =\sum_i c^i_4  \,  I^{(i)}_4\,,
\end{equation}
where  $\Phi=q$  for the  fermionic  loop or  $\Phi=\varphi$ for  the
complex scalar loop.

The above result is expressed in  terms of scalar box integral
functions evaluated in $D=4-2\epsilon$ that
carry $\epsilon$ singularities.
Because of manifest one-loop ultra-violet and infra-red
finiteness   of  the   one-loop  amplitude both  the   $1/\epsilon^2$  and
$1/\epsilon$  poles  must  cancel  between  the various  terms  in  the
amplitude. Because  the amplitude reduces to  scalar box contributions
it is immediate (but tedious) to evaluate the expression for the amplitude
from its quadruple cut. It would very be interesting to understand this
generic structure  valid for all  helicity configurations from  a dual
Wilson loop description as in~\cite{Alday:2007hr,Drummond:2007aua}.
This in turn implies relations between the box coefficients which
will be discussed in the subsequent section.

\subsection{Universal ultra-violet and infra-red pole structure}
The infra-red singularities in an ordered gauge theory one-loop amplitude
is described by~\cite{Giele:1991vf}
\begin{equation}\label{e:IRsafe}
  \cA^\ol_n(k_1,\dots,k_n)       \Big|_{IR}       \propto      \sum_{i=1}^n
  {(-(k_i+k_{i+1})^2)^{-\epsilon}\over \epsilon^2}\,
  \cA^\tree_n(k_1,\dots,k_n)\,.
\end{equation}
Since the $n$-photon tree-level  amplitude are vanishing in an Abelian
theory like QED, there is  no infra-red singularities in the $n$-photon
amplitudes     at     one-loop.      As     a     result,     previous
calculations~\cite{Mahlon:1993fe,Jikia:1993tc,Gounaris:1999gh,Mahlon:1994dc,Binoth:2007ca,Bernicot:2007hs}
were  expressed in  terms of  the finite  part $\mathcal{F}_4$  of the
scalar  box  integral   function  $I_4$.  This is because  the  dimensionless
one-mass and  two-mass triangle integral  functions are given  by (see
the appendix~\ref{sec:scalarBox})
\begin{eqnarray}
  \tilde I_3(k_1,k_2,K_3) &\equiv& (-K_3^2)\,I_3(k_1,k_2,K_3) ={ r_\Gamma\,\mu^{2\epsilon}\over
    \epsilon^2} \,(-K_3^2)^\epsilon\,, \\
\nn \tilde I_3(k_1,K_2,K_3) &\equiv& (K_2^2-K_3^2)\,I_3(k_1,K_2,K_3) ={ r_\Gamma\,\mu^{2\epsilon}\over
    \epsilon^2} \,\left((-K_2^2)^{-\epsilon}-(-K_3^2)^{-\epsilon}\right)\,,
\end{eqnarray}
and we can define the finite parts $\mathcal{F}_4$ of
the scalar box
integral functions $I_4$ by subtracting dimensionless one-mass and two-mass divergent
scalar triangle functions
\begin{equation}
  \mathcal{F}_4= I_4+ \sum_{i} t^1_i \, \tilde I^{(i)\,\rm 1-mass}_3+
  \sum_{i} t^2_i \, \tilde I^{(i)\,\rm 2-mass}_3\,.
\end{equation}
Here $t^1_i$ and $t^2_i$ are some coefficients depending on the kinematic
invariants which are given in the appendix~\ref{sec:scalarBox}.
Choosing this basis makes the amplitude explicitly free from divergences but
hides the
no-triangle  property  given by equation~\eqref{e:Ampn8}.
Because the  dimensionless one-mass triangle  gives the multi-particle
infra-red  divergence $(-K^2)^{-\epsilon}/\epsilon^2$,  the  absence of
triangles lead to a set of relations between the box coefficients appearing in the
decomposition~(\ref{e:Ampn8}). In the next
section we  will give  for the specific  example of  MHV multi-photon
amplitudes the relations
between the box coefficients for the cancellation of the infra-red divergences.

We would like to contrast this to the gravity case where the infra-red
singularities are given
by~\cite{Dunbar:1995ed}
\begin{equation}
\cM^\ol_n(k_1,\dots,k_n)       \Big|_{IR}       \propto    \cM^\tree_n(k_1,\dots,k_n) \,  \sum_{i=1}^n
{(-(k_i+k_{i+1})^2)^{1-\epsilon}\over \epsilon^2}\,.
\end{equation}
The     leading    $1/\epsilon^2$     pole    in   the  gravity
amplitude cancels~\cite{GreenSW}    because   of    the on-shell
condition $\sum_{i=1}^n (k_i+k_{i+1})^2=0$ but the amplitude has
still a non-vanishing $1/\epsilon$ pole contribution in
$D=4-2\epsilon$.

By  power counting,  ultra-violet divergences  can only  occur  in
the three-photon  and  four-photon  one-loop amplitudes.  The
three-photon amplitude vanishes by Furry's theorem while the
four-photon amplitudes are non-vanishing.   The  four-photon
amplitude at one-loop  is  dimensionless  in  four  dimensions, and
could  have a  logarithmic ultra-violet  divergence.  However  such
an ultra-violet divergence has to be associated with a local gauge
invariant operator $T_{m_1n_1\cdots m_4n_4}F^1_{m_1n_1}\cdots
F^4_{m_4n_4}$, for which the four photons amplitude is given by some
combination of the four field-strengths
$F_{mn}=\epsilon_mk_n-\epsilon_nk_m$ of the external photons.
Because such an operator has mass dimension four, no ultra-violet
divergences can occur by dimensional analysis.
Therefore all multi-photon one-loop amplitudes are
ultra-violet finite and all possible rational pieces
contributions  are of infra-red
origin.  The presence of rational contributions  will  be analysed
in section~\ref{sec:rationalterms} following the method
of~\cite{Forde:2007mi,Badger:2008cm}.

\subsubsection{The $n$-photon MHV Amplitude}

In this section we re-evaluate    the    $n$-photon
one-loop    MHV    amplitude
$A^\ol_{n;q}(k_1^-,k_2^-,\break k_3^+,\dots,k_n^+)$ for $n\geq8$. This
amplitude was first computed by Mahlon in~\cite{Mahlon:1994dc} and has been recently re-analysed
using double unitarity cuts~\cite{Bernicot:2007hs}. We present it again here in order to analyse the
infra-red structure of the $n\geq8$-photon MHV amplitudes which have only box contributions.

Because  of the restrictions  on the  cut momenta~\cite{Britto:2004nc}
these MHV amplitudes are only given by the linear combination of
the one-mass box $I^{1m}_4(k_i^+,k_{a}^-,k_j^+,k_b^-+K^+_1)$  with
the massless  legs  given by  the configurations
$(k_i^+,k_a^-,k_j^+)$  with   $3\leq  i<j\leq  n$  and $a,b=1,2$,
and the massive leg $K^+_1= k_3+\cdots +k_{n}-k_i-k_j$. The
two-mass easy box  $I^{2me}_4(k_i^+,k_a^++K_2^+,k_j^+,k_b^-+K_3^+)$
with the opposite  massless legs is given by the configuration
$(k_i^+,k_j^+)$ with $3\leq i<j\leq n$ and the two massive legs
$k_a^-+K^+_2$ and $k_b^-+K^+_3$ with $a,b=1,2$ and $K^+_2+K^+_3=
k_3+\cdots +k_{n}-k_i-k_j$.
Because the cut amplitude only involves MHV tree-amplitude factors we can make
use of the compact formula of eq.~(\ref{e:KSMHV}) for the tree amplitudes
in the cut leading to
\begin{equation}
c_{4;k_i|k_1+K_1|k_j|k_2+K_2}=\AB{k_i}{K_1^\flat}{k_i}\AB{k_j}{K_1^\flat}{k_j}
\times \tilde c_{4;k_i|k_1|k_j|k_2}\,,
\end{equation}
with
\begin{equation}
  K_1^\flat= K_1- {K_1^2\over \spab{k_1}.{K_1}.{k_i}}\, k_i\,,
\end{equation}
and
\begin{equation}
 \tilde c_{4;k_i|k_1|k_j|k_2}= {1\over2}
\left({\A{k_1}{k_i}^3\A{k_1}{k_j}^2\A{k_2}{k_i}^2\A{k_2}{k_j}^3\over
    \A{k_i}{k_j}^8 }+(1\leftrightarrow 2)\right)\, \prod_{1\leq r\leq n\atop r\neq i,j}{\A{k_i}{k_j}\over \A{k_i}{k_r}\A{k_r}{k_j}}\,.
\end{equation}
Here $K_1+K_2$ is a repartition of the positive helicity states on
each of the opposite corners of the box.
The coefficient $\tilde c_{4;k_i|k_1|k_j|k_2}$ does  not  depend on
the distribution of  the positive helicity states of  the opposite
massive legs and  gives  the same expression for the one-mass
box and  the  two-mass  easy  box.
This is compatible with the soft limit relation between the
two-mass easy box and the one-mass box
\begin{equation}
  \lim_{K_2\to0} I^{2me}_4(k_i,k_a+K_1,k_j,k_b+K_2)= I^{1m}_4(k_i,k_a+K_1,k_j,k_b)\,.
\end{equation}
Under the exchanges of the two positive helicity
massless legs the two-mass-easy box coefficient has the parity
\begin{equation}
  c_{4,k_i|k_1|k_j|k_2}= (-1)^n \, c_{4,k_j|k_1|k_i|k_2}\,.
\end{equation}
This implies that the only amplitude with an even number of
external photons lines  is non-vanishing. This  is a particular
example  of the consequence of Furry's theorem on the coefficients
of the scalar box integrals.

Using that
\begin{equation}
\AB{k_i}{K_1^\flat}{k_i}\AB{k_j}{K_1^\flat}{k_j} =  (k_i+k_1+K_1)^2
(k_j+k_1+K_1)^2 - (k_1+K_1)^2 (k_2+K_2)^2\,,
\end{equation}
one can express the one-loop amplitude in terms of the dimensionless
boxes $   \tilde     I^{2me}_4(k_1,K_2,k_3,\break K_4)=
(s_{12}s_{23}-K_2^2K_4^2)\,  I_4^{2me}(k_1,K_2,k_3,K_4)$

\begin{equation}\label{e:MHV2}
  A^\ol_{2n;q}(k_1^-,k_2^-,k_3^+,\cdots,k_{2n}^+)=  \sum_{3\leq
    i\neq                j\leq                2n}              \tilde
  c_{4;k_i|k_1|k_j|k_2} \sum_{
    K_1^{(ij)}} \tilde I^{2me}_4(k_i,k_1+K_1^{(ij)},k_j,k_2+K_2^{(ij)})\,,
\end{equation}
where we have made use of the notation  $K_1^{(i_1\cdots i_r)}$
defined to be the sum of external momenta such that $K_1^{(i_1\cdots
i_r)}+K_2^{(i_1\cdots i_r)}= k_3+\cdots +k_{2n}-k_{i_1}-\cdots
-k_{i_r}$.
Using the $\epsilon$ expansion given in the appendix~\ref{sec:scalarBox} and
the symmetry of the coefficient in the exchange between $i$ and $j$, we
can isolate the infra-red divergent part of this amplitude
\begin{eqnarray}
A^\ol_{2n;q}(k_1^-,k_2^-,k_3^+,\cdots,k_{2n}^+)\Big|_{IR}&=&  {2r_\Gamma\,\mu^{2\epsilon}\over\epsilon^2}      \sum_{3\leq
    i\neq j\leq 2n\atop  K_1^{(ij)}} \!\!\!\tilde
  c_{4;k_i|k_1|k_j|k_2}\times\\
\nn &\times&
  \, \left((-(k_1+k_i+K_1^{(ij)})^2)^{-\epsilon}- (-(k_1+K_1^{(ij)})^2)^{-\epsilon}\right)\,.
\end{eqnarray}
The infra-red  divergence associated with  the multi-particle invariant
$(k_1+K_1^{(ij)})^2$ is given by
\begin{eqnarray}
&&\!\!\!\!\!\!\!\!\!\!\!\!\!\!\!\!\!\!\!\!\!\!\!\sum_{i,j=3\atop
i\neq j}^{2n}\!\!
        c_{4;k_i|k_1\!+\!K_1^{(ij)}|k_j|k_2\!+\!K_2^{(ij)}}
        \!+\!2\!\!\sum_{l=3\atop l\neq i,j}^{2n}\!\!
         c_{4;k_i|k_1\!+\!K_1^{(ijl)}|k_l|k_2\!+\!K_2^{(ijl)}}
        \!+\!\sum_{l=3\atop r\neq l}^{2n}\!
        c_{4;k_l|k_1\!+\!K_1^{(lr)}|k_r|k_2\!+\!K_2^{(lr)}}
=0\,,
\end{eqnarray}
which shows that  the amplitude is free of  infra-red divergences, as it
should be, since all the soft factors are vanishing for a multi-photon amplitude.

\section{Conclusions}
In  this  paper  we  have  considered amplitudes  in  unordered  field
theories such  as gravity and  QED. New integral  reduction formul\ae\
derived using the world-line formalism  have been examined and we have
seen how such formul\ae\ can have a wide range of applications in four
dimensional theories.

It was  shown in~\cite{BjerrumBohr:2008ji,BjerrumBohr:2008dp} that, for
maximal  $\cN=8$ supergravity,  the constraints  from the  new integral
reduction formul\ae\ leads to the `no-triangle' property for $n$-point
supergravity amplitudes.   In cases with less  supersymmetry {\it e.g.}
$\cN=4$ supergravity it means that the $n$-graviton amplitude contains
only integral functions  up to scalar bubble integrals  and that it is
constructible           from           its           cuts           in
$D=4-2\epsilon$~\cite{Bern:2007xj,BjerrumBohr:2008ji,BjerrumBohr:2008dp}. For
pure gravity our result yields  an amplitude consisting of scalar box,
triangle and  bubble integrals as  well as rational  polynomial terms.
These results are completely  surprising from na{\"\i}ve power counting
arguments.

In this paper we have showed that one can apply the reduction formul\ae\
eq.~(\ref{e:Red1}) and eq.~(\ref{e:Red2}) to the one-loop multi-photon
amplitudes in QED to  obtain that the one-loop multi-photon amplitudes
with at least eight external  photons are given by scalar box integral
functions only.   Such amplitudes satisfy a  no-triangle property from
$n\geq  8$ and  are thus  given solely  by their  quadruple  cut.  The
amplitudes  contain no  rational polynomial  contributions.   This `no
triangle'  property  of  multi-photon  QED amplitude  with  $n\geq  8$
photons is  true for helicity  configurations of the  external photons
generalising  the pure MHV  results of  Mahlon~\cite{Mahlon:1993fe}. This
result  is clearly  unexpected  from  na{\"\i}ve power  counting
arguments.   Of  course  the  appearance  of the  various  scalar  box
integral functions in the expression  for the amplitude depends on the
helicity  configuration   of  the   external  states.   It   would  be
interesting   to  reproduce   these  results   for   generic  helicity
configurations using a Wilson loop representation of the amplitude~\cite{Alday:2007hr,Drummond:2007aua}.

We have shown  that the considered cancellations can  be made manifest
by  a  choice   of  transverse  gauge  and  the   summation  over  the
permutations  of  the  unordered  legs.   We  expect  that  unexpected
cancellations should also appear in amplitudes with mixed photon-gluon
external states.   For such amplitudes  one should expect a  number of
cancellations in the summation over the unordered photon lines.

Investigations   of  higher   loop  multi-photon   and  multi-graviton
amplitudes   presents  another   interesting  direction   for  further
investigation.  Within the unitarity method formalism the cancellations
seen for  one-loop unordered  amplitudes pose various  restrictions on
the  type  of  integral   functions  appearing  in  the  expansion  of
multi-loop  amplitudes~\cite{Bern:2006kd}.    Factorisation  based  on
`no-triangle'   properties  at   one-loop   definitely  suggest   that
amplitudes should  have a simpler  form (due to  cancellations between
orderings) than  na{\"\i}ve counting  proposes.  For maximal  ${\cal N}
=8$  supergravity   this  gives  a  necessary   (but  not  sufficient)
requirement  for the  absence  of the  three-loop  divergence in  four
dimensions~\cite{Bern:2007hh}.

The  results  of this  paper  show that  a  world-line  approach is  a
particularly  good  framework for  analysing  the  properties of  loop
amplitudes  in   unordered  field  theories.   An   extension  of  the
world-line formalism  to higher  loops would be  very helpful  in this
respect  and would  be  required  for a  better  understanding of  the
perturbative  structure  of  $\cN=8$  supergravity.  This  would  help
justifying    and    resolving    the   various    constraints    from
supersymmetry~\cite{Berkovits:2006vc}                               and
dualities~\cite{Green:2006gt,Green:2006yu}  in   four  dimensions  and
might  lead  to a  conclusive  argument  for  or against  perturbative
finiteness of maximal ${\cal N} =8$ supergravity in $D=4$.

\section*{Acknowledgements}

We would like to thank Zvi Bern, David Kosower and Lance Dixon for many
enlightening discussions.  We are grateful to Lance Dixon for attracting our
attention to the case of QED amplitudes.  We would like to thank  as well
Zoltan Kunszt and Pierpaolo Mastrolia for comments on the draft and Gregory
Korchemsky for discussions. We also thanks Rutger Boels for bringing to our
attention an error in a previous version of this manuscript. This research was
supported in part (NEJBB) by grant DE-FG0290ER40542 of the US Department of
Energy and (PV) by the RTN contracts MRTN-CT-2004-503369, MRTN-CT-2004-005104,
as well  as the ``Agence Nationale de la Recherche'' grants BLAN06-3-137168 and
ANR-05-BLAN-0073-01. SB  also acknowledges support  from the Helmholtz
Gemeinschaft under contract VH-NG-105.

\appendix

\section{Helicity formalism conventions}\label{sec:conventions}
All conventions and the notation in the paper follows that of
ref.~\cite{Mangano:1990by} unless otherwise stated.

We will here employ the mostly minus metric signature
$\eta^{\mu\nu}=\textrm{diag}(+,-,-,-)$ and use a representation of
the Dirac matrices satisfying
$\{\gamma^\mu,\gamma^\nu\}=2\eta^{\mu\nu}$, {\it i.e.},
\begin{equation}
\gamma^\mu =
\begin{pmatrix}
  0 & \sigma^\mu \\
  \bar{\sigma}^\mu & 0
\end{pmatrix}; \qquad
\gamma_5=\begin{pmatrix}
1 & 0
\\ 0 & -1
\end{pmatrix}\,.
\end{equation}
Here $(\sigma^\mu)=(1,\sigma^i)$ and
$(\bar\sigma^\mu)=(-1,\bar\sigma^i)$ and $\sigma^i$ are the standard
Pauli matrices. We will make use of the slashed notation $\gamma^\mu
p_\mu=\sla p$.

For any light-like momentum $p$ the positive energy solution to the
Dirac equation is $\sla p  \, u_h(p)=0$ both for positive and
negative helicities, {\it i.e.}, $h=+1$ and $h=-1$. This solution
satisfy the chirality condition $(1\mp \gamma_5)/2\, u_\pm(p)=0$ and
$(1\pm\gamma_5)/2\,\bar u_\pm(p)=0$.

We will make use of the following conventions
\begin{eqnarray}
|k\rangle{}&\equiv u_+(k);\qquad |k]{}&\equiv
u_-(k) \\
\langle k|{}&\equiv \bar u_-(k);\qquad {}[k|{}&\equiv \bar u_+(k)\,.
\end{eqnarray}
Spinor products will be defined according to
\begin{equation}
\an[p,q]{} \equiv\bar u_-(p) u_+(q);\qquad \sq[p,q]{}\equiv \bar
u_+(p) u_-(q)\,,
\end{equation}
where $(p+q)^2=2p\cdot q=\an[p,q]\sq[p,q]$.

With these conventions the completeness relation gives that
\begin{equation}
\sum_{h=\pm1} u_h(k)\bar u_h(k)=\sla k= |k \rangle[k|+ |k]\langle k|\,.
\end{equation}
The polarisation tensor for the photon of light-like momentum $k$ can be
represented as
\begin{equation}
\label{e:pol} 
\epsilon^+_\mu(k,p_{\rm ref})
= { [ k | \gamma_\mu | p_{\rm ref} \ra
\over 
\sqrt2 \an[p_{\rm ref},k]};
\qquad 
\sla \epsilon^-_\mu(k,p_{\rm ref})
= -{ \la k | \gamma_\mu | p_{\rm ref} ]
\over 
\sqrt2 \sq[p_{\rm ref},k]}
\,,
\end{equation}
where $p_{\rm ref}$ is an arbitrary light-like reference momentum.

\section{On the improved large $z$ behaviour of the $e^-e^+\to n\, \gamma$
tree-amplitude}\label{sec:proof}

In this section we will discuss the large $z$ behaviour of the $e^+e^-\to n\gamma$ tree-level amplitude
under the BCFW shifts, see~(\ref{e:S1}) and~(\ref{e:S2}). The
behaviour of the scalar amplitude can be related to that of the
fermion by supersymmetric Ward identities.

We will first write the amplitude with $n$ external photons in the
following way
\begin{equation}
A^\tree_{n;q}\big(p_a^h,p_b^{-h};k_1,\dots,k_n\big)=
\sum_{i=1}^n\,\sum_{\sigma\in\mathfrak{S}_{n-1}}
(-1)^{n-i}      {      e\over         n}\, \bar u_{h}(p_a)
\mathfrak{J}^\tree_{1\cdots \hat\imath\cdots n}\, {\sla p_b+\sla
k_i\over            (p_b+k_{i})^2}\, \sla \epsilon_{i}
u_{-h}(p_b)\,. \label{e:step1}
\end{equation}
Where $\mathfrak{J}^\tree_{1\cdots
\hat \imath \cdots n}$
is an off-shell current constructed from the remaining $n-1$ photons
after the $i^{\rm th}$ photon is removed from the list. We have
\begin{equation}
  \label{e:As}
  \mathfrak{J}^\tree_{1\cdots \hat \imath \cdots n}\equiv
  \sla \epsilon_{1} {i
    \sla q_1\over    q_{1}^2}  \cdots \sla \epsilon_{i-1}    {i   \sla q_{i-1}\over
    q_{i-1}^2} \sla \epsilon_{i+1} {i \sla q_{i+1}\over
    q_{i+1}^2}\cdots {i \sla q_{n-1}\over
    q_{n-1}^2} \sla \epsilon_{n}\,.
\end{equation}
Here $q_j=p_a+k_1+\cdots +k_j$ where, as before, we have not included the momentum of the $i^{\rm th}$
state. Using that $\sla   p_b \sla \epsilon_i+  \sla \epsilon_i \sla p_b=2\,p_b\cdot \epsilon_i$ and
$\sla p_b \,u_{-h}(p_b)=0$ we rewrite
\begin{equation}
\sla  p_b\sla \epsilon_i \,u_{-h}(p_b)= 2\, u_{-h}(p_b)\, p_b\cdot
\epsilon_i\,.
\end{equation}

Choosing the reference momentum of the photons to be $p_{\rm
ref}=p_b$ so that $\epsilon_i\cdot p_b=0$ one can rewrite the
tree-amplitude as
\begin{equation}
A^\tree_{n;q}\big(p_a^h,p_b^{-h};k_1,\dots,k_n\big)=
\sum_{i=1}^n\,\sum_{\sigma\in\mathfrak{S}_{n-1}}
(-1)^{n-i}      {      e\over         n}\, {\bar u_{h}(p_a)
\mathfrak{J}^\tree_{1\cdots \hat \imath \cdots n}\sla k_{i}\, \sla
\epsilon_{i} u_{-h}(p_b)\over            (p_b+k_{i})^2}\,.
\label{e:step2}
\end{equation}
With this choice of reference momentum we also have that
\begin{equation}
\sla \epsilon_i^- \,u_-(p_b)=0, \qquad \sla
\epsilon_i^+\,u_+(p_b)=0\,.
\end{equation}
Only the non-zero contributions are such that the helicity of the
$i^{\rm th}$ photon is the opposite of the one of the positron. We remark
as well that
\begin{equation}
{\sla k_i            \sla \epsilon_i^h           \,u_{-h}(p_b)\over
(p_b+k_i)^2}={\sla k_i\,u_{-h}(p_b)\over\sqrt2
(p_b+k_i)^2}={u_{-h}(k_i)\over \sqrt2\langle
k_i^{h}|p_b^{-h}\rangle}\,.
\end{equation}
Combining these properties we arrive at the following expression for the tree-level amplitude
\begin{equation}
A^\tree_{n;q}\big(p_a^h,p_b^{-h};k_1,\dots,k_n\big)=
\sum_{i=1}^n\,\sum_{\sigma\in\mathfrak{S}_{n-1}}
(-1)^{n-i}    \,\delta(h_i+h=0)\,  {      e\over         n}\, {\bar
u_{h}(p_a) \mathfrak{J}^\tree_{1\cdots \hat \imath \cdots n}\,
u_{-h}(k_i)\over      \sqrt2   \langle k_i^{-h}|p_b^{h}\rangle}\,.
\label{e:current}
\end{equation}
This means that  the   multi-photon  tree-level  amplitude has been
rewritten as a sum of $(n-1)$-photon \emph{off-shell} currents,
$\mathfrak{J}^\tree_{1\cdots \hat \imath \cdots n}$. All external photons in this expression have $p_b$
as their reference momentum. We can now study the large $z$ behaviour of the tree-level amplitude
$A^\tree_{n;q}$ under the BCFW shifts~(\ref{e:S1}) and~(\ref{e:S2}).

Because the reference momentum of the photons is $p_b$ we have that
for each polarisation choice, in the limit where ${z\to\infty}$,
that the polarisation tensor behave as
\begin{eqnarray}\label{e:limP}
 \sla\epsilon^+_\infty(k,p_b)=\lim_{z\to\infty}        \sla       \epsilon^+(k,\widehat       p_b)=
 \lim_{z\to\infty}{|\widehat p_b\rangle [k|\over \an[\widehat p_b,k]}=
 {s+1\over2 } {|p_b\rangle [k|\over \an[p_b,k]} + {s-1\over2}
 {|p_a\rangle [k|\over \an[p_a,k]}\,,
\end{eqnarray}
and
\begin{eqnarray}\label{e:limN}
\sla\epsilon^-_\infty(k,p_b)= \lim_{z\to\infty} \sla
\epsilon^-(k,\widehat p_b)= \lim_{z\to\infty}{|\widehat
   p_b]\langle k|\over \sq[\widehat p_b,k]}=
 {s+1\over2   }  {|p_a]\langle   k|\over  \sq[p_a,k]}   +  {s-1\over2}
 {|p_b]\langle k|\over \sq[p_b,k]}\,,
\end{eqnarray}
which is independent of $z$.

We will now consider the behaviour of the product of a fermion propagator
and a polarisation
\begin{equation}
T_{i}= \sla\epsilon^\pm(k_i,\widehat p_b){\sla {\widehat q}_i\over
{\widehat q}_i^2}\,.
\end{equation}
Under the BCFW shift the momentum $\sla{\widehat q}_i$ shift according
eq.~(\ref{e:qshift})
\begin{equation}
\sla{\widehat q}_i = \sla q_i + z\sla \pi\,,
\end{equation}
where $\sla\pi$ is defined in eq.~(\ref{e:pi})

\begin{equation}
\sla \pi \equiv  {1+s\over2} \,\Big( |  p_b\rangle
[p_a| +  | p_a] \langle p_b| \Big) +{s-1\over2}\, \Big( |
p_a\rangle [p_b| + | p_b] \langle p_a|\Big)\,.
\end{equation}
The rest of the discussion we can choose $s=1$ for   illustration.
We can then show that, (for $ h=\pm1$)
\begin{eqnarray}
\nn  
\lim_{z\to\infty} \sla \epsilon^{h_i}(k_i,\widehat p_b)\sla{\widehat
  q}_i&=& 
  \sla \epsilon^\pm_\infty(k_{i}, p_b)\,\sla q _i\,
  + z \sla K_i\\
\sla K_i&=& \sqrt2\,
\begin{pmatrix}
   {h_i-1\over2}\, {\an[k_i,p_b]\over\sq[k_i,p_a]}\,|p_a][p_a|& 0\\ 0 &
   {h_i+1\over2}\,{\sq[k_i,p_a]\over\an[k_i,p_b]}\, |p_b\rangle \langle p_b|
  \end{pmatrix}
\, .
\end{eqnarray}
%
%
%
Since 
\begin{equation}\label{e:Kidentity}
K^h\,K^{h'} = \bar{u}(p_a)\,\sla\! K^{h=-1} =
  \sla\! K^{h=1} v(p_b) = 0\,,
\end{equation} 
the terms with two consecutive powers
  of $K^h$ will always vanish in each current.
We introduce the quantity $\Delta_i= z\, \sla\!K_i/\wh{q}_i^2$
  such that  $\tilde T_i=z\, (T_i - \Delta_i)$ has the large $z$ behaviour:
\begin{eqnarray}
 \tilde T_i^\infty= \lim_{z\to\infty} \tilde T_i  = {\sla \epsilon_\infty^\pm(k_i,p_b)\,\sla
  q _i \over 2(q_i\cdot\pi)}\,.
\end{eqnarray}
We now consider the two ends of the amplitude $A_{n;q}^\tree$ written in the form of
\eqref{e:current}
\begin{equation}
  T_0 = \tilde{T}_0 = \bar u_h(\widehat p_a) ,\qquad
  T_{n} = \tilde{T}_n = {\sla\epsilon^\pm(k_n,\wh{p}_b) u_{-h}(k_i)\over \langle k_i^{-h}|\widehat
  p_b^h\rangle}\,.
\end{equation}
Clearly $T_n$ has the large $z$ behaviour given by
\begin{equation}
  \lim_{z\to\infty} T_n = \lim_{z\to\infty} \, {\sla\epsilon^\pm(k_n,\wh{p}_b) u_{-h}(k_i)\over \langle
k_{i}^{-2h}|p_b^{2h}\rangle}\sim z^{hs-1/2}\,
\sla\epsilon^\pm_\infty(k_n,p_b)u_{-h}(k_i){\langle
k_{i}^{h}|p_b^{-h}\rangle \over 2\, \pi\cdot k_i}; \qquad
h=\pm\frac12\,.
\end{equation}
For $T_0$ (by definition of the shift on the fermion polarisations) we have
\begin{equation}
  \lim_{z\to\infty} T_0= z^{hs+1/2}\, T_0^\infty; \qquad
h=\pm\frac12\,.
\end{equation}
Each term in the sum~(\ref{e:current})  
\begin{equation}
\lim_{z\to\infty}{\bar{u}_h(p_a)\mathfrak{J}^\tree_{1\cdots \hat \imath \cdots n}u_{-h}(k_i) \over \langle k_{i}^{-2h}|p_b^{2h}\rangle}
=\lim_{z\to\infty} T_0T_1\cdots T_{i-1}T_{i+1}\cdots T_{n-1}T_n\,,
\end{equation}
are expanded according the number of $\Delta$ insertions
\begin{eqnarray}\label{e:sumTtilde}
\lim_{z\to\infty}\, z^{-2h+n-2}\,{\bar{u}_h(p_a)\mathfrak{J}^\tree_{1\cdots \hat \imath \cdots n}u_{-h}(k_i) \over \langle k_{i}^{-2h}|p_b^{2h}\rangle}
&=& \tilde T_0^\infty\tilde{T}^\infty_1\cdots
\tilde{T}^\infty_{i-1}\tilde{T}^\infty_{i+1}\cdots \tilde{T}^\infty_{n-1}\tilde{T}^\infty_n\\
+z\sum_{j=1\atop j\neq i}^{n-1} \tilde T_0 \tilde T_1 \cdots
\Delta^{h_j}  \cdots \tilde
T_{n-1}\tilde T_n
&+&\cdots+z^{\lceil {n-2\over2}\rceil} \tilde T_0 \tilde T_1\Delta^{h_2}\tilde T_3 \cdots \Delta^{h_{n-1}}\tilde T_n\nonumber\,,
\end{eqnarray}
according to~\eqref{e:Kidentity}. The worst large $z$ behaviour, $z^{2h+1-\lfloor{n-2\over2}\rfloor}$, is obtained for the maximal
  number of $\Delta$ insertions.  Compared to the na\"\i ve  behaviour,
  $z^{2h}$, 
  we already see an improvement for each terms in~(\ref{e:current})
\emph{before} performing the sum over the permutations.

Summing over the permutations the 
  contributions with at least one $\Delta^h$ insertion cancel and the large $z$ scaling property of the $e^+e^-\to n\,\gamma$ tree-level amplitude
under the BCFW shifts~(\ref{e:S1}) and~(\ref{e:S2}) is given by the
first line of~\eqref{e:sumTtilde}.  We have checked this numerically
for up to $n=7$ photon lines. The numerical analysis indicates that for MHV
amplitudes only the first term in~\eqref{e:sumTtilde} gives a non
vanishing contribution, therefore providing a direct indication for the 
scaling properties of the amplitudes at large $z$.

The analysis for the fermion case in QED is rather special because
the fermions can only be adjacent in the interactions that are
involved. This is however {\it not} the case for generic amplitudes
for example in gravity. As well it is the absence of photon
  self-interactions that allow  the
sum over the permutations  to improve the large $z$ behaviour.
Such cancellations are not expected in generic QCD amplitudes  because of the gluon
self interactions.

Using supersymmetric  Ward identities  we can conclude that the
massless scalar tree amplitudes have the large $z$ scaling given by
\begin{equation}
  \lim_{z\to\infty} A_{n;\varphi}^\tree\sim {1\over z^{n-2}}\,.
\end{equation}
In the scalar  case as  explained in section~\ref{sec:sumdisc} a
gauge choice is not enough for obtaining this behaviour and extra
cancellations has to arise in the sum over orderings.

\section{Cut analysis of one-loop $n$-photon integral coefficients~\label{app:cutanalysis}}

In this appendix we will give further details regarding the
tree-level $z$-scaling, eq.~\eqref{e:largez} of the amplitude. The
knowledge of the tree amplitudes BCFW $z$-scaling behaviour combined
with an analysis of unitarity cuts for example using a formalism
such as Forde~\cite{Forde:2007mi} have been used successfully to
show analogous simplifications in gravity theories see
refs.~\cite{BjerrumBohr:2006yw,Bern:2007xj}. For demonstrating the
absence of scalar triangle and bubble integrals in the one-loop
$n$-photon amplitude for $n\geq6$ this is a very powerful strategy.
To examine the analytic structure of the rational terms we use
$D$-dimensional cutting
techniques~\cite{Bern:1995db,Bern:1996ja,Anastasiou:2006gt,Britto:2008vq,Giele:2008ve}.
We will prove the vanishing of rational polynomial terms in the QED
amplitudes using a generalisation of Forde's method for
$D$-dimensional cuts, re-expressed in terms of massive
four-dimensional cuts~\cite{Badger:2008cm}.

\subsection{Absence of triangles in amplitudes with $n>6$}\label{sec:triangle}
\begin{figure}[ht]
\begin{center}
\includegraphics[width=5cm]{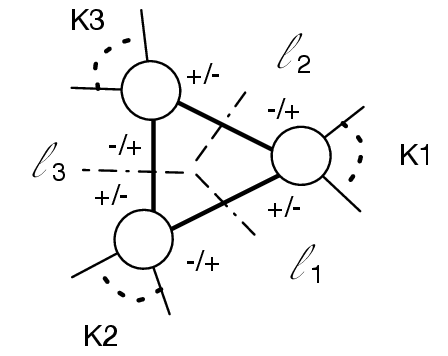}
 \caption{\label{fig:triplecut} The triple-cut contribution the amplitude.}
\end{center}
\end{figure}

We will first analyse the presence of triangles in the $n$-photon
one-loop amplitude, by considering the triple cut represented in
figure~\ref{fig:cut}(b)
\begin{equation}
\label{e:3cut} \left.      A^\ol_{n;q} \right|_{ \textrm{3-cut}} \!\!\!=
(2\pi)^3\!\!\int  \!   d^4\ell \!    \prod_{i=1}^3\delta(\ell_i^2)
\sum_{h=\pm\frac12}\!\!
A^\tree_{n_1}(\ell_1^{2h},-\ell_2^{-2h})A^\tree_{n_2}(\ell_2^{2h},-\ell_3^{-2h})
A^\tree_{n_3}(\ell_3^{2h},-\ell_1^{-2h})\,.
\end{equation}
Here  $n_1+n_2+n_3=n$.  We have only indicated the momenta of the
fermions, which are given by the cut loop momenta.
Following~\cite{Forde:2007mi}  we can  parametrise  the  loop
momenta $\ell_i$ with $i=1,2,3$ in the cut as
\begin{equation}
  \sla \ell_i= t\,|K_1^{\flat}]\langle K_2^{\flat}|+{\alpha_{i1}\alpha_{i2}\over
    t}|K_1^{\flat}\rangle [ K_2^{\flat}|+ \alpha_{i1}\sla K_2^{\flat}+\alpha_{i2}\sla
    K_1^{\flat}\,,
\end{equation}
where $\ell=\ell_3$. The shifted propagator factors behaves as
\begin{equation}
  \lim_{t\to\infty}           \widehat  {\sla     q}_i\sim       t
  |K_1^{\flat}]\langle K_2^{\flat}|\,.
\end{equation}
With this parametrisation of the loop momenta the polarisations of
the fermions will be given by
\begin{eqnarray}
  \label{e:li3cut}
u_-(\ell_i)&\equiv&|\ell_i] =t| K_1^{\flat}] +\alpha_{i1} |
  K_2^{\flat}]\\
\nn u_+(\ell_i)&\equiv& | \ell_i\rangle ={\alpha_{i2}\over t}|
K_1^{\flat}\rangle +|  K_2^{\flat}\rangle\,,
\end{eqnarray}
which have  the large $t$ behaviour
\begin{equation}
 \lim_{t\to\infty} u_-(\ell_i)\sim t\, |K_1^{\flat}],\qquad
 \lim_{t\to\infty} u_+(\ell_i)\sim |K_2^{\flat}\rangle\,.
\end{equation}
The large $t$ scaling  is equivalent to the  BCFW shift scaling of
type $s=+1$ in~(\ref{e:S1})  (except for  the  scaling of the
$u_-(\ell_i)$ which scales like $z$ in~(\ref{e:ushift})). Because of
this  the amplitude will  have only less power  of $t$ in the
numerator for the external fermion line. Of course for the scalar
tree-level amplitude there is no $t$ factor from the external scalar
states.

Taking into account the scaling of the external states
we can conclude (using the found large $z$  scaling of
the  tree amplitudes given in~(\ref{e:largez})) that the large
$t$ behaviour is
\begin{eqnarray}
  \label{e:larget}
  \lim_{t\to\infty}             A^\tree_n(\ell_i^{2h},-\ell_j^{-2h},k_1,\dots,k_n)\sim
 t^{2-n}\,.
\end{eqnarray}
We remark  that this behaviour is  independent of the  helicity $h$ of
the state running in the loop. Therefore the
large $t$ behaviour of the cut~(\ref{e:3cut}) is given by
\begin{equation}\label{e:larget2}
 \lim_{t\to\infty}
 A^\tree_{n_1}(\ell_1^{2h},-\ell_2^{-2h})A^\tree_{n_2}(\ell_2^{2h},-\ell_3^{-2h})A^\tree_{n_3}(\ell_3^{2h},-\ell_1^{-2h})\sim
 {1\over t^{n-6}}\,.
\end{equation}
We can thus conclude that the one-loop $n>6$ photon amplitude with
either a fermion or a scalar running in the loop  do not contain
scalar triangle integrals with massive corners.

\subsection{Absence of bubbles in amplitudes with $n>4$}\label{sec:bubble}
\begin{figure}[ht]
\begin{center}
\includegraphics[width=9cm]{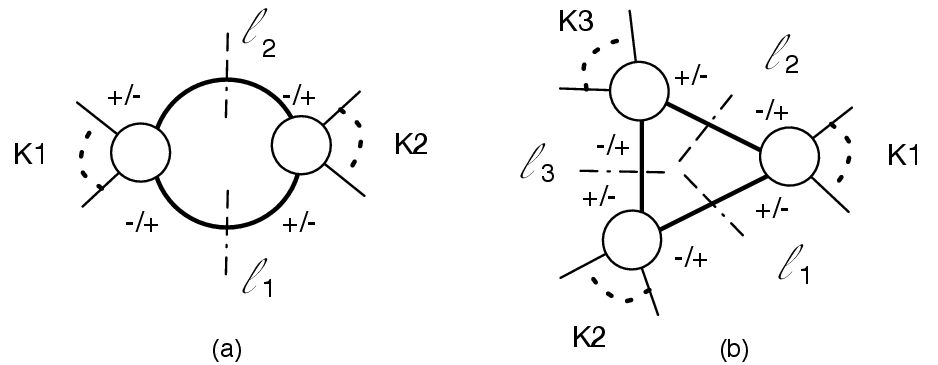}
 \caption{\label{fig:doublecut} The double-cut contribution to the amplitude
 is given by (a)
   the bubble cut and (b) the triangle subtraction.}
\end{center}
\end{figure}
In   this    section   we   examine   the   two-line    cut   of
the amplitude.

The cut amplitude can be computed as
\begin{equation}\label{e:2cut}
\left.      A^\ol_{n;q} \right|_{ \textrm{2-cut}} = (2\pi)^3\int
d^4\ell_1     \, \delta(\ell_1^2)\delta(\ell_2^2)
\;\sum_{h=\pm\frac12} A^\tree_{n_1}(\ell_1^{2h},-\ell_2^{-2h}) \,
A^\tree_{n_2}(\ell_2^{2h},-\ell_1^{-2h})\,,
\end{equation}
with $n_1+n_2=n$. The cut loop momenta can be parametrised as
follows~\cite{Forde:2007mi}
\begin{equation}
  | l_1]=  t |  K_1^{\flat}]  +(1-y)  \,  {K_1^2\over
    \gamma}\,|  \chi^-],  \qquad  |  l_1\rangle={y\over  t}\,
  | K_1^{\flat}\rangle+|\chi\rangle\,,
\end{equation}
and
\begin{equation}
 |  l_2]= |  K_1^{\flat}] -{y\over  t}\, {K_1^2\over
  \gamma}|   \chi],\qquad   |  l_2\rangle=(y-1)\,   |
K_1^{\flat}\rangle+t | \chi\rangle\,.
\end{equation}
For large $t$ with $t\gg y$,  the behaviour is like the one of the
triangle cut analysed  in the  previous section. The  analysis in
the previous section showed that no triangles are present in
amplitudes with $n\geq 7$  photons, so  for these  amplitudes  the
leading  behaviour of  the amplitude   for  $y\gg   t\gg   1$   will
be   a   test  for possible  bubble contributions.  For  $n=6$  the
large $t$  behaviour  of  the  triple-cut amplitude is  given by a
constant. In that case as  well the leading behaviour  of the
amplitude  for $y\gg  t\gg  1$ will  be enough  for analysing the
bubble contributions. For $n\leq 5$ one has to pay attention to the
triangle subtractions~\cite{Forde:2007mi} represented in
fig.~\ref{fig:doublecut} that can lead to a contribution in  the
regime  where $y\sim t$.   We will  discuss these contributions
below.

For analysing the  presence of scalar bubble integrals one needs to
take the limits $y\to \infty$, $t\to\infty$ and $y\gg t\gg 1$. In
this case the polarisation of the fermion shifts according to
\begin{eqnarray}
  \label{e:li2cut}
 u_-(\ell_1)&\sim -y\, \, {K_1^2\over
    \gamma}\,|  \chi],\qquad
u_+(\ell_1)&\sim {y\over t}| K_1^{\flat}\rangle\,,\\
\nn u_-(\ell_2)&\sim-{y\over t}| K_1^{\flat}],\qquad
  u_+(\ell_2)&\sim y\,| K_1^{\flat}\rangle\,.
\end{eqnarray}
The polarisations of the scalar fields do not shift. The shifted
propagator factors behaves as
\begin{equation}
  \lim_{t\to\infty}   \widehat{\sla  q}_i\sim -{y^2\over t}\, {K_1^2\over
    2\gamma}\,
  |K_1^{\flat}]\langle \chi|\,.
\end{equation}
From the  analysis of  the large  $z$ behaviour of  the BCFW  shift
in section~\ref{sec:BCFW} we  conclude that the  large $y\gg t \gg
1$ of the tree amplitudes behave as
\begin{eqnarray}
   A^\tree_{n_1}(\ell_1^{-2h},-\ell_2^{+2h})&\sim& \left(y^2\over
    t\right)^{2-n_1}\,t^{2h}\,,\\
\nn  A^\tree_{n_2}(\ell_2^{-2h},-\ell_1^{+2h})&\sim&
\left(y^2\over
    t\right)^{2-n_2}\, t^{-2h}\,,
\end{eqnarray}
for $h=\pm\frac12$ for the fermion loop and $h=0$ for the scalar
loop. And the integrand of the cut amplitude in~(\ref{e:2cut})
behaves as
\begin{equation}\label{e:bubbles}
  A^\tree_{n_1}(\ell_1^{2h},-\ell_2^{-2h})A^\tree_{n_2}(\ell_2^{2h},-\ell_1^{-2h})\sim  \left(y^2\over
    t\right)^{4-n}\,.
\end{equation}
The  answer  is independent  of  the helicity and of the nature of
the particle (fermionic or scalar) running in the loop.

From this scaling  we can conclude that no  bubbles appear in
amplitudes with $n\geq6$  photons. Furthermore the amplitude with
four photons contains bubble  contributions as directly confirmed by
the explicit evaluation of the
amplitude~\cite{Jikia:1993tc,Gounaris:1999gh}.

For $n=5$ photons   the leading  behaviour in~(\ref{e:bubbles})
vanishes and no   pure scalar bubble  contributions   are  found in
the amplitude. In this case there are non-vanishing
 triangle subtractions from the  regime where $y\sim t$ as represented in
 fig.~\ref{fig:doublecut}.  All amplitudes  vanishes via Furry's
 theorems and the triangles contributions add up to zero via
 symmetry  properties of  the
 amplitude~(\ref{e:Symm}) and after summing
 over  the orderings of external legs.

\subsection{Absence of rational terms}\label{sec:rationalterms}

In this section we connect  the large $z$
behaviour~(\ref{e:largezmassive}) of the massive scalar
tree-amplitude to the large $\mu^2$ limit of massive scalar loop
amplitudes for testing for rational terms contributions
following the method used
in~\cite{Forde:2007mi,Badger:2008cm}. In  this section we  will
follow closely the notations and conventions
of ref.~\cite{Badger:2008cm}. The analysis makes use of the $D$-dimensional integral basis recently used
in numerical implementations of the $D$-dimensional unitarity method \cite{Giele:2008ve}.

To analyse the rational contributions to the $n$-photon loop
amplitudes we need to consider $D$-dimensional unitarity cuts. We
can relate the $D$-dimensional loop momenta to massive momentum
using:
\begin{equation}
    \ell_{[D]}^\nu = \bl^\nu+\ell_{[-2\e]}^\nu,\qquad
    \bl^2 = \mu^2\,.
\end{equation}
Using a $D$-dimensional integral basis it is possible to write the
rational contributions in a basis of massive box, triangle and
bubble functions by relating the $D$-dimensional integral to
integrals of the mass parameter, $\mu$. This results in \cite{Giele:2008ve}:
\begin{align}
    R_n =
    \sum_{K_4}\, C^{[4]}_{4;K_4}I_{4;K_4}[\mu^4]+\sum_{K_3}\,      C^{[2]}_{3;K_3}      I_{3;K_3}[\mu^2]+\sum_{K_2} \,
    C^{[2]}_{2;K_2}I_{2;K_2}[\mu^2]\,,
\end{align}
where $K_i$  with $1\leq i  \leq 4$ denote  the set of momenta  of
the massive scalar box integral functions.

After performing the $\bl\sim \mu$ loop momentum integration and
taking the $\e\to0$ limit this becomes~\cite{Giele:2008ve}:
\begin{align}
R_n =
-\frac{1}{6}\sum_{K_4}C^{[4]}_{4;K_4}-\frac{1}{2}\sum_{K_3}C^{[2]}_{3;K_3}
-\frac{1}{6}\sum_{K_2}C^{[2]}_{2;K_2}\,K_2^2\,.
\end{align}
The coefficients $C^{[4]}_{4;K_4},C^{[2]}_{3;K_3}$ and
$C^{[2]}_{2;K_2}$ can then be extracted from the analysis of the
large momentum scaling of the generalised cuts with four dimensional
massive propagators.

For multi-photon  loop amplitudes the rational terms  can be
extracted from tree amplitudes  with a massive fermion in the  loop.
The use of  massive fermions  is however  delicate  to use in
amplitudes evaluated  in $D=4-2\epsilon$.   Therefore   we    will
use   the supersymmetric decomposition to  write, the QED  amplitude
as the one-loop amplitude for $\cN=1$ super-QED minus the
contributions of a scalar loop
\begin{equation}
    A_{n;q}^\ol= A_{n;{{\cal N}=1}}^\ol-A_{n;\varphi}^\ol.
\end{equation}
Since  any supersymmetric  amplitudes  are cut  constructible in
four dimensions,  all  the rational  pieces  are arising  from  the
scalar  loop contribution. The rational part contribution to a
scalar amplitude can be extracted by  introducing an effective mass
$\mu^2$  for the scalar and by evaluating  the integral coefficient
in four dimensions  with the tree-level amplitudes for massive
scalars~\cite{Badger:2005zh,Forde:2005ue,Brandhuber:2005jw}.

\subsubsection*{$\mu^2$ Dependence Of The Box Coefficients}

For  the  quadruple  cut  of  the  massive  scalar  loop
multi-photon amplitude we choose the following basis for the loop
momentum (see section~4.1 of \cite{Badger:2008cm} for notation),
\begin{equation}
    {\sla \bl}_1 = a \skf4 + b \skf1 + c|\kf4\ra[\kf1| +
    \frac{\gamma_{14}\, ab-\mu^2}{\gamma_{14}\, c}|\kf1\ra[\kf4|\,,
\end{equation}
where the on-shell constraints determine
\begin{eqnarray}
  \gamma_{14}&=& K_1\cdot K_4\pm \sqrt{K_1\cdot K_4- K_1^2\,K_4^2}\,,\\
  a&=&{K_1^2(K_4^2+\gamma_{14})\over \gamma_{14}^2-K_1^2K_4^2};\qquad
  b={K_4^2(K_1^2+\gamma_{14})\over \gamma_{14}^2-K_1^2K_4^2}\,,
\end{eqnarray}
which do  not depend  on $\mu$, and  two solutions for  $c=c_\pm$ that
have the large $\mu$ limit
\begin{equation}
\lim_{\mu^2\to\infty}   c_\pm =
    \pm|\mu|\sqrt{\frac{\AB{\kf1}{K_2}{\kf4}}{\gamma_{14}\,\AB{\kf4}{K_2}{\kf1}}}\,.
\end{equation}
The box-type rational contribution is given by the large $\mu$
behaviour of the quadruple cut:
\begin{equation}\label{e:C44}
    C^{[4]}_4 =
    \frac{i}{2}\sum_{c=c\pm}\inf_{\mu^2}[A^\tree_{n_1}A^\tree_{n_2}A^\tree_{n_3}A^\tree_{n_4}(\bl_1(c))]\Big|_{\mu^4}\,,
\end{equation}
with $n_1+n_2+n_3+n_4=n$. For  a  function   $f(x)$  with  at  most
a   polynomial  growth  for $x\to\infty$
\begin{equation}
  \lim_{x\to\infty} f(x) = a_n x^n +\cdots +a_0+ \mathcal{O}(1/x)\,,
\end{equation}
we define the `$\inf$' operation following~\cite{Berger:2006ci}
\begin{equation}
  \inf_x f=  a_n x^n +\cdots +a_0\,.
\end{equation}
We will use  as well the notation $\inf_x  f\Big|_{x^k}= a_k$ for
the coefficient of $x^k$.

To see the cancellation of such  terms in the photon amplitudes we
use the  scaling of a  generic tree amplitude  in the regime  of large
$\mu$ derived in section~\ref{sec:massive}.

For large $\mu$, since $c\sim \mu$ the loop momenta in the cuts
scale as
\begin{equation}
  \lim_{\mu^2\to\infty}\,  \bl_1(c)\sim \mu\, \chi\,,
\end{equation}
where $\chi$ is  some non-vanishing vector, which is  the behaviour of
eq.~(\ref{e:largepa}), and the  analysis   of   the  end   of
section~\ref{sec:massive} gives that the
massive scalar tree-level amplitude has the behaviour

\begin{equation}
\lim_{\mu^2\to\infty} A^\tree_{n} = {1\over \mu^{n-2}}\,.
    \label{eq:treescale}
\end{equation}
This  implies   that  the  product  of  the   four  tree-level
factor in~(\ref{e:C44}) has the large $\mu$ behaviour
\begin{equation}
  \lim_{\mu^2\to\infty}                                              \,
  A^\tree_{n_1}A^\tree_{n_2}A^\tree_{n_3}A^\tree_{n_4}(\bl_1(c))\sim
  {\mu^4\over \mu^{n-4}}\,,
\end{equation}
implying that $C_4^{[4]}(k_1,\cdots,k_4)$ does not vanish for the
four-photon  amplitudes which hence will receive a  contribution
from rational pieces     in    agreement     with    the explicit
computation in~\cite{Mahlon:1993fe,Jikia:1993tc,Gounaris:1999gh}.
For more  than four photons  the one-loop amplitude  does not have
any rational term contribution
\begin{equation}
    C_4^{[4]}(k_1,\ldots,k_n) = 0 \qquad \text{for } n>4
    \label{eq:R4vanish}\,.
\end{equation}

\subsubsection*{\bf $\mu^2$ Dependence Of The Triangle Coefficients}

For the triple  cut of the massive scalar  loop multi-photon
amplitude we choose the  following basis for the loop  momentum (see
section~4.2 of~\cite{Badger:2008cm})
\begin{equation}
    \sla \bl_1 = a\kf4 + b\kf1 + t|\kf4\ra[\kf1| +
    \frac{\gamma_{14}ab-\mu^2}{\gamma_{14}t}|\kf1\ra[\kf4|\,,
    \label{eq:3cutloopmom}
\end{equation}
with the same expressions for $\gamma_{14}$  and $a$ and $b$ as in
the previous section. In general there will be two solutions to the
on-shell constraints $\bl_1^2=0$ which we label $\bl_1^\pm$. We have
\begin{equation}
    C^{[2]}_3                                                     =
        \frac{1}{2}\sum_{\sigma=\pm}\inf_{\mu^2}[\inf_t[A^\tree_{n_1}A^\tree_{n_2}A^\tree_{n_3}(\bl_1^\sigma)]]\Big|_{t^0,\mu^2}\,,
\end{equation}
with $n_1+n_2+n_3=n$. We must consider the product of three tree
amplitudes in  the $t\gg\mu\to \infty$  limit. In this limit  the loop
momenta takes the following asymptotic form
\begin{equation}\label{e:asymp}
\lim_{\mu^2\to\infty}\lim_{t\to\infty}     \sla    \bl_1     \sim t
|\kf4\ra[\kf1| - {\mu^2\over t}\,{1\over \gamma_{14}}\,
|\kf1\ra[\kf4|\,.
\end{equation}
Following     the    analysis     of    the     triangle     cut in
section~\ref{sec:triangle}  we  deduce  that in this limit  the
massive scalar tree amplitude scale like
\begin{equation}
  \label{e:scalem}
  \lim_{t\to \infty} A^\tree_{n;\varphi}\sim  {1\over t^{n-2}}
  \,      \left(C^\infty_{n;\varphi}+      {\mu^2\over      t^2}\,\delta
    C^\infty_{n;\varphi} + \mathcal{O}(\mu^4/t^4)\right)\,.
\end{equation}
The  sub-leading  corrections in  $\mu^2$  arises  from the
$\mu^2/t$ dependence in~(\ref{e:asymp}) and because $\mu^2$ is
dimensionful these corrections appear with  a factor of order
$1/\bl^2$ so that  the corrections are of order
$\mathcal{O}((\mu^2/t)\times 1/\bl^2)\sim  \mathcal{O}(\mu^2/ t^2)$
in the large $t$ limit where $t\gg \mu^2$.

For $n>6$ the product of the tree amplitude hence behave as
$O(1/t^{n-6})$ therefore
\begin{equation}
  \inf_t[
  A^\tree_{n_1}A^\tree_{n_2}A^\tree_{n_3}(\bl_1^\sigma)]=0\,,
\end{equation}
so we can conclude that $C^{[2]}_3=0$ for the one-loop amplitude
with $n>6$ external photons.
For $n=6$ we have for each triangle contribution
\begin{eqnarray}
  \inf_t[C_3^{N_3-\rm mass}]\Big|_{t^0}&=&{1\over2}\sum_{\sigma=\pm}\,
  C^\infty_{n_1;\varphi}
  C^\infty_{n_2;\varphi} C^\infty_{n_2;\varphi}(\bl_1^\sigma); \qquad
  N_3=1,2,3\,.
\end{eqnarray}
The leading $\mu^2$ contribution does  not depend on $\mu^2$ and
there is no rational term contribution from scalar triangle
integrals for the $n=6$ photon amplitude. The sub-leading
contributions to the tree-amplitude are    of     order
$\mathcal{O}(\mu^2/t^{n-1})$. This imply that  the contribution to
the triple cut has the large $t$ expansion
\begin{equation}
  \inf_t[C_3^{N_3-\rm mass}]=\inf_t\Big[{\mu^2\over t^2} \delta C^\infty_{n_1;\varphi}
  C^\infty_{n_2;\varphi} C^\infty_{n_3;\varphi}\Big]
=0; \qquad N_3=1,2,3\,,
\end{equation}
and hence there is no rational term contribution from triangles for
$n=6$ photons.

Only the  $n=4$ photon  amplitude can  get a  contribution from the
one-mass triangle.  The product of the tree amplitudes leads to a
$\mu^2$ contribution

\begin{eqnarray}\label{e:C31m}
    \inf_t[C_3^{\rm                     1-mass}]\Big|_{t^0}&=&\inf_t\Big[
  t^2\, \prod_{i=1}^3 (C^\infty_{n_i;\varphi}+{\mu^2\over                      t^2}\delta
    C^\infty_{n_i;\varphi})\Big]\Big|_{t=0}\\
\nn&=&\mu^2( C^\infty_{n_1;\varphi}C^\infty_{n_2;\varphi}\delta
C^\infty_{n_3;\varphi}+ {\rm perm.}(n_1,n_2,n_3))\,.
\end{eqnarray}
These coefficients can be given by the sum over the permutations of
the corresponding gluon amplitude evaluated
in~\cite{Bern:1995db,Britto:2008vq,Badger:2008cm}.

\subsubsection*{\bf $\mu^2$ Dependence Of The Bubble Coefficients}

For the double cut of the massive scalar loop multi-photon
amplitude we choose the following basis for the loop  momentum (see
section~4.3 of~\cite{Badger:2008cm})

\begin{equation}
    \sla \bl_1 = y\kf1 + \frac{K_1^2(1-y)}{2(K_1\cdot \chi)}\XX + t|\kf1\ra[\XX_1| +
    \frac{y(1-y)K^2_1-\mu^2}{2t\,(K_1\cdot \chi)}|\XX\ra[\kf1|\,,
\end{equation}
with $K_1^\flat =K_1 -\chi\, K_1^2/(2(K_1\cdot \chi))$.
The bubble coefficient has two components, a pure double cut term
and a set of triangle subtraction terms:
\begin{align}
    C_2^{[2]} = C_2^{ {\rm bub},[2]} + \sum_{\{K_3\}}C_2^{ {\rm tri}(K_3)[2]}\,.
    \label{eq:ddbub}
\end{align}
These components are expressed in terms of the large momentum
scaling as,
\begin{align}
 \label{e:C2b}   C_2^{           {\rm           bub}          [2]}           &=
        -i\inf_{\mu^2}[\inf_t[\inf_y[A_{n_1}^\tree A_{n_2}^\tree(\bl_1(y,t,\mu^2)]]]\Big|_{\mu^2,t^0,y^i\to
    Y_i},\\
\label{e:C2t}    C_2^{ {\rm tri}(K_3) [2]} &=
    -\frac{1}{2}\sum_{\sigma=\pm}
    \inf_{\mu^2}[\inf_t[A_{n_1}^\tree A_{n_2}^\tree A_{n_3}^\tree
    (\bl_1^\sigma(y_\sigma,t,\mu^2)]]\Big|_{\mu^2,t^i\to T_i}\,.
\end{align}
The non-vanishing  integrals    depend on $\mu^2$ and  have the
following large $\mu^2$  behaviour (see section~4.3
of~\cite{Badger:2008cm} for detailed expressions)
\begin{eqnarray}
   \lim_{\mu^2\to\infty} Y_0 &=& \mathcal{O}(1)\,, \quad
   \lim_{\mu^2\to\infty}  Y_1 = \mathcal{O}(1)\,, \quad
   \lim_{\mu^2\to\infty} Y_2 =\mathcal{O}(\mu^2)\,,
    \label{eq:Ys}\\
    \label{eq:Ts}
    \lim_{\mu^2\to\infty} T_1 &=& \mathcal{O}(1)\,, \quad
    \lim_{\mu^2\to\infty} T_2 = \mathcal{O}(1)\,, \quad
    \lim_{\mu^2\to\infty} T_3 = \mathcal{O}(\mu^2)\,.
\end{eqnarray}

With    an    analysis   similar    to    the    one   performed in
section~\ref{sec:bubble} we obtain that the product of the tree
amplitudes in~(\ref{e:C2b}) and~(\ref{e:C2t}) behaves as
$O((y^2/t)^{4-n})$ therefore for $n>4$ external photons we have
\begin{equation}
  \inf_t[\inf_y[ A_{n_1}^\tree A_{n_2}^\tree(\bl_1)]] =0\,.
\end{equation}
Hence there  is no rational term  contributions from the  bubbles to
the one-loop amplitude with  $n>4$ external photons. For $n=5$ there
is {\it a priori}  a  non   vanishing  contribution  from the
subtraction of triangles $C_2^{{\rm tri}[K3]}$  but as before  these
contributions  vanish in  the sum over all the orderings as required
by Furry's theorem.

Both   $C_2^{\rm  bub[2]}$ and $C_3^{{\rm tri}(K_3)[2]}$ for four
external photons receive non-zero contributions which can be
obtained by summing over   the ordering of the  corresponding  gluon
amplitude contributions which were evaluated
in~\cite{Bern:1995db,Britto:2008vq,Badger:2008cm}.

Therefore  there is a rational term contribution to the four point
amplitude in agreement with the explicit computations performed
in~\cite{Mahlon:1993fe,Jikia:1993tc,Gounaris:1999gh}.

\section{The scalar box integral functions}\label{sec:scalarBox}

In this section we give a relation between the infra-red part of the
massless scalar box integral  functions   evaluated  in
$D=4-2\epsilon$ dimensions  and triangle contributions. We will
follow the notation of the refs.~\cite{Bern:1993kr,Ellis:2007qk}.

We will use  $k_i$ for massless legs $k_i^2=0$,  and $K_i$ for massive
legs  $K_i^2\neq0$.   As  well  we   will  use  $s_{ij}$   for  either
$-(k_i+k_j)^2$, or $-(k_i+K_j)^2$, or $-(K_i+K_j)^2$.

The infra-red divergent part of  the massless scalar box integral
function is given by
\begin{equation}
  \label{e:massless}
  I_4(k_1,k_2,k_3,k_4)\Big|_{IR}=         r_\Gamma\,      {\mu^{2\epsilon}\over
    s_{12}s_{23}}\, {2\over\epsilon^2} \,\left((-s_{12})^{-\epsilon}+(-s_{23})^{-\epsilon}\right)\,.
\end{equation}
The infra-red divergent part of  the one-mass scalar box integral
function is given by
\begin{equation}
  \label{e:1mass}
  I_4(k_1,k_2,k_3,K_4)\Big|_{IR}=r_\Gamma\,
        {\mu^{2\epsilon}\over
    s_{12}s_{23}}\, {2\over\epsilon^2} \,\left((-s_{12})^{-\epsilon}+(-s_{23})^{-\epsilon}-(-K_4^2)^{-\epsilon}\right)\,.
\end{equation}
The infra-red divergent part of  the two-mass easy scalar box
integral function is given by
\begin{equation}
  \label{e:2masseasy}
 I_4(k_1,K_2,k_3,K_4)\Big|_{IR}=   r_\Gamma
        {\mu^{2\epsilon}\over
    s_{12}s_{23}-K_2^2K_4^2} {2\over\epsilon^2} \left((-s_{12})^{-\epsilon}\!+\!(-s_{23})^{-\epsilon}
    \!-\!(-K_2^2)^{-\epsilon}\!-\!(-K_4^2)^{-\epsilon}\right)\,.
\end{equation}
The infra-red divergent part of  the two-mass hard scalar box
integral function is given by
\begin{eqnarray}
  \label{e:2masshard}
\nn  I_4(k_1,k_2,K_3,K_4)\Big|_{IR}&=&   r_\Gamma\,
        {\mu^{2\epsilon}\over
    s_{12}s_{23}}\,                                  {2\over\epsilon^2}
  \,\left((-s_{12})^{-\epsilon}+(-s_{23})^{-\epsilon}-(-K_3^2)^{-\epsilon}-(-K_4^2)^{-\epsilon}\right)\\
\nn&+&r_\Gamma {\mu^{2\epsilon}\over
    s_{12}s_{23}}\,                {1\over\epsilon^2}               \,
  {(-K_3^2)^{-\epsilon}(-K_4^2)^{-\epsilon}\over
    (-s_{12})^{-\epsilon}}\\
\nn &=&r_\Gamma\, {\mu^{2\epsilon}\over
    s_{12}s_{23}}\,                                  {1\over\epsilon^2}
  \,\left((-s_{12})^{-\epsilon}+2(-s_{23})^{-\epsilon}-(-K_3^2)^{-\epsilon}-(-K_4^2)^{-\epsilon}\right)\,.\\
\end{eqnarray}
The infra-red divergent part of  the three-mass scalar box integral
function is given by
\begin{eqnarray}
  \label{e:3mass}
\nn   I_4(k_1,K_2,K_3,K_4)\Big|_{IR}&=&   r_\Gamma\,
        {\mu^{2\epsilon}\over
    s_{12}s_{23}-K_2^2K_4^2}\,                       {2\over\epsilon^2}
  \,\left((-s_{12})^{-\epsilon}+(-s_{23})^{-\epsilon}\right)\\
\nn&-&   r_\Gamma\,
        {\mu^{2\epsilon}\over
    s_{12}s_{23}-K_2^2K_4^2}\,{2\over\epsilon^2}\, \left((-K_2^2)^{-\epsilon}+(-K_3^2)^{-\epsilon}+(-K_4^2)^{-\epsilon}\right)\\
\nn&+&  r_\Gamma\,{\mu^{2\epsilon}\over
    s_{12}s_{23}-K_2^2K_4^2}\,                {1\over\epsilon^2}               \,
 \Big(                   {(-K_2^2)^{-\epsilon}(-K_3^2)^{-\epsilon}\over
   (-s_{23})^{-\epsilon}}+
 {(-K_3^2)^{-\epsilon}(-K_4^2)^{-\epsilon}\over
   (-s_{12})^{-\epsilon}}\Big)\\
\nn&=&r_\Gamma\,  {\mu^{2\epsilon}\over
    s_{12}s_{23}-K_2^2K_4^2}\,                       {1\over\epsilon^2}
  \,\left((-s_{12})^{-\epsilon}+(-s_{23})^{-\epsilon}-(-K_2^2)^{-\epsilon}-(-K_4^2)^{-\epsilon}\right)\,,\\
\end{eqnarray}
where
$r_\Gamma=\Gamma(1+\epsilon)\Gamma(1-\epsilon)^2/\Gamma(1-2\epsilon)$.

The divergent dimensionless one-mass and two-mass scalar triangle functions are given by
\begin{eqnarray}
  \label{e:TriDiv}
 \tilde I_3(k_1,k_2,K_3)&\equiv& (-K_3^2)\, I_3(k_1,k_2,K_3)= r_\Gamma\,{\mu^{2\epsilon} \over\epsilon^2}\,
  (-K_3^2)^{-\epsilon}\,,\\
 \tilde I_3(k_1,K_2,K_3)&\equiv& (K_2^2-K_3^2)\, I_3(k_1,k_2,K_3)=r_\Gamma\,{\mu^{2\epsilon}      \over\epsilon^2}\,
  \left((-K_2^2)^{-\epsilon}-(-K_3^2)^{-\epsilon}\right)\,.
\end{eqnarray}
These  expressions   imply  that  all   the  divergent  parts   of
the dimensionless scalar box integral functions can be  expressed as
linear  combination of  the infra-red  parts of  the dimensionless
scalar triangle functions in the following way
\begin{eqnarray}
\tilde I_4(k_1,k_2,k_3,k_4)\Big|_{IR}&=&2\, \Big( \tilde
I_3(k_3,k_4,k_1+k_2)+ \tilde I_3(k_1,k_4,k_2+k_3)\Big)\Big|_{IR}\,,\\
\tilde I_4(k_1,k_2,k_3,K_4)\Big|_{IR}&=&
\Big(\tilde I_3(k_1,k_2,k_3+K_4)+\tilde I_3(k_2,k_3,k_1+K_4)\\
\nn&+& \tilde I_3(k_3,k_1+k_2,K_4)+
    \tilde I_3(k_1,k_2+k_3,K_4)\Big)\Big|_{IR}\,,\\
\tilde I_4(k_1,K_2,k_3,K_4)\Big|_{IR}&=& 2 \,
\Big(\tilde I_3(k_1,K_2,k_3+K_4)
+ \tilde I_3(k_1,k_3+K_2,K_4)\Big)\Big|_{IR}\,,\\
\tilde I_4(k_1,k_2,K_3,K_4)\Big|_{IR}&=&
\Big(\tilde I_3(k_1,k_2,K_3+K_4)+    \tilde
I_3(k_1,k_2+K_3,K_4)\\
\nn&+& \tilde I_3(k_2,K_3,k_1+K_4)\Big)\Big|_{IR}\,,\\
\tilde I_4(k_1,K_2,K_3,K_4)\Big|_{IR}&=&
\Big(\tilde I_3(k_1,K_2,K_3+K_4)
+\tilde I_3(k_1,K_2+K_3,K_4)\Big)\Big|_{IR}\,.
\end{eqnarray}

\end{document}